\documentclass[article]{jss-alt}

\usepackage{thumbpdf,lmodern}
\usepackage{amsmath,amssymb,amsfonts,amsthm,bbm,mathrsfs,mathrsfs,dsfont}
\usepackage{subfig}
\usepackage{lscape}

	\renewcommand{\P}{\mathbb{P}}
	\newcommand{\I}{\mathds{1}}
	
  \DeclareMathOperator*{\argmin}{arg\,min}

  \newcommand{\Trans}{\mathsf{T}}
  
\newcommand{\Bb}{\mathbf{b}}
\newcommand{\Be}{\mathbf{e}}
\newcommand{\Br}{\mathbf{r}}

\newcommand{\BW}{\mathbf{W}}
\newcommand{\BX}{\mathbf{X}}
\newcommand{\BY}{\mathbf{Y}}

\newcommand{\Bbeta}{\boldsymbol{\beta}}

\usepackage{framed}

\author{
Matias D. Cattaneo \\ Princeton University
\And 
Michael Jansson \\ UC Berkeley \\ \textit{CREATES}
\And 
Xinwei Ma \\ UC San Diego
}
\Plainauthor{Matias D. Cattaneo, Michael Jansson, Xinwei Ma}

\title{\pkg{lpdensity}: Local Polynomial Density Estimation and Inference}
\Plaintitle{lpdensity: Local Polynomial Density Estimation and Inference}
\Shorttitle{\pkg{lpdensity}: Local Polynomial Density Estimation and Inference}

\Abstract{
Density estimation and inference methods are widely used in empirical work. When the underlying distribution has compact support, conventional kernel-based density estimators are no longer consistent near or at the boundary because of their well-known boundary bias. Alternative smoothing methods are available to handle boundary points in density estimation, but they all require additional tuning parameter choices or other typically \textit{ad hoc} modifications depending on the evaluation point and/or approach considered. This article discusses the \proglang{R} and \proglang{Stata} package \pkg{lpdensity} implementing a novel local polynomial density estimator proposed and studied in \citet*{Cattaneo-Jansson-Ma_2020_JASA,Cattaneo-Jansson-Ma_2021_LocRegDistribution}, which is boundary adaptive and involves only one tuning parameter. The methods implemented also cover local polynomial estimation of the cumulative distribution function and density derivatives. In addition to point estimation and graphical procedures, the package offers consistent variance estimators, mean squared error optimal bandwidth selection, robust bias-corrected inference, and confidence bands construction, among other features. A comparison with other density estimation packages available in \proglang{R} using a Monte Carlo experiment is provided.
}

\Plainkeywords{kernel-based nonparametrics, local polynomial, density estimation, bandwidth selection, bias correction, robust inference, boundary carpentry, R, Stata}

\Address{
Matias D. Cattaneo\\
Department of Operations Research and Financial Engineering\\
Princeton University\\
227 Sherrerd Hall\\
Princeton, New Jersey 08544, USA\\
E-mail: \email{cattaneo@princeton.edu}\\
URL: \url{https://cattaneo.princeton.edu/}\\

Michael Jansson\\
Department of Economics\\
University of California, Berkeley\\
530 Evans Hall \#3880\\
Berkeley, California 94720, USA\\
E-mail: \email{mjansson@econ.berkeley.edu}\\
URL: \url{https://eml.berkeley.edu/~mjansson/}\\

Xinwei Ma\\
Department of Economics\\
University of California, San Diego\\
9500 Gilman Dr. \#0508\\
La Jolla, California 92093, USA\\
E-mail: \email{x1ma@ucsd.edu}\\
URL: \url{https://sites.google.com/view/xinweima/}
}

\begin{document}

%% -- Introduction -------------------------------------------------------------

%% - In principle "as usual".
%% - But should typically have some discussion of both _software_ and _methods_.
%% - Use \proglang{}, \pkg{}, and \code{} markup throughout the manuscript.
%% - If such markup is in (sub)section titles, a plain text version has to be
%%   added as well.
%% - All software mentioned should be properly \cite-d.
%% - All abbreviations should be introduced.
%% - Unless the expansions of abbreviations are proper names (like "Journal
%%   of Statistical Software" above) they should be in sentence case (like
%%   "generalized linear models" below).

%%%%%%%%%%%%%%%%%%%%%%%%%%%%%%%%%%%%%%%%%%%%%%%%%
%% SECTION: Introduction                       %%
%%%%%%%%%%%%%%%%%%%%%%%%%%%%%%%%%%%%%%%%%%%%%%%%%
\section{Introduction}\label{sec:intro}

Nonparametric estimation of a probability density function (PDF), as well as its associated cumulative distribution function (CDF) or higher-order derivatives thereof, plays an important role in empirical work across many disciplines. Sometimes these quantities are the main objects of interest, while in other cases they are useful ingredients in forming more complex nonparametric or semiparametric statistical procedures. See \citet{Wand-Jones_1995_Book} and \citet{Fan-Gijbels_1996_Book} for classical textbook introductions to kernel-based density and local polynomial methods.

This article discusses the main methodological and numerical features of the software package \pkg{lpdensity}, available in both \proglang{R} and \proglang{Stata}, which implements the local polynomial smoothing approach proposed and studied in \citet*{Cattaneo-Jansson-Ma_2020_JASA,Cattaneo-Jansson-Ma_2021_LocRegDistribution} for estimation of and inference on a smooth CDF, PDF, and derivatives thereof. In a nutshell, the idea underlying this estimation approach is to first approximate the discontinuous empirical CDF using local polynomial methods, and then employ that smoothed approximation to construct estimators of the distribution function, density function, and higher-order derivatives.

The resulting local polynomial density estimator is intuitive and easy to implement, and exhibits several interesting theoretical and practical features. For example, it does not require pre-binning or any other complicated pre-processing of the data, and enjoys all of the celebrated features associated with local polynomial regression estimation \citep{Fan-Gijbels_1996_Book}. In particular, it automatically adapts to the (possibly unknown) boundaries of the density's support, a feature that is unavailable for most other density estimators in the literature. See \cite{Karunamuni-Albert_2005_SM} for a review on this topic. Two exceptions are the local polynomial density estimators of \cite*{Cheng-Fan-Marron_1997_AoS} and \cite{Zhang-Karunamuni_1998_JSPI}, which require pre-binning of the data or, more generally, pre-estimation of the density near the boundary, thereby introducing additional tuning parameters that need to be chosen for implementation. In contrast, the density estimator implemented in the \pkg{lpdensity} package requires choosing only one tuning parameter: the bandwidth entering the local polynomial approximation, for which the package also offers data-driven selectors. Furthermore, following the results in \citet*{Calonico-Cattaneo-Farrell_2018_JASA,Calonico-Cattaneo-Farrell_2020_CEOptimal}, robust bias-corrected inference methods are also implemented, which allow using mean squared error (MSE) optimal or the integrated mean squared error (IMSE) optimal bandwidth choices when forming confidence intervals or conducting hypothesis testing.

The software implementation covers smooth estimation of the distribution and density function, and derivatives thereof, for any polynomial order at both interior and boundary points. \citet*{Cattaneo-Jansson-Ma_2020_JASA,Cattaneo-Jansson-Ma_2021_LocRegDistribution} give formal large-sample statistical results for these estimators, including (i) asymptotic expansions of the leading bias and variance, (ii) asymptotic pointwise and uniform normal approximations, (iii) consistent standard error estimators, (iv) consistent data-driven bandwidth selection based on asymptotic MSE expansions of the point estimators, and (v) asymptotically valid uniform inference and confidence bands. Importantly, all these results apply to both interior and boundary points simultaneously. We briefly summarize these results in the upcoming sections, and illustrate them numerically, including a comparison with other methods available in \proglang{R}.

In the remaining of this article, we focus on the \proglang{R} implementation of the software package \pkg{lpdensity} (\url{https://CRAN.R-project.org/package=lpdensity}), but all functionalities are also available in \proglang{Stata}. See Appendix \ref{appendix:stata} for more details. The \proglang{R} package includes the following two main functions.
\begin{itemize}
\item \code{lpdensity()}. This function implements the local polynomial approximation to the empirical distribution function for a grid of evaluation points, and offers smooth point estimators of the CDF, PDF, and derivatives thereof. The function takes the bandwidth for each evaluation point as given, and employs the companion function \code{lpbwdensity()} for data-driven bandwidth selection whenever the bandwidth is not provided. Inference is implemented by using robust bias correction methods \citep*{Calonico-Cattaneo-Farrell_2018_JASA,Calonico-Cattaneo-Farrell_2020_CEOptimal}, and both pointwise confidence intervals and uniform confidence bands are supported. Standard inference methods assuming undersmoothing or ignoring smoothing bias are also available. 

\item \code{lpbwdensity()}. This function offers pointwise and integrated MSE-optimal bandwidth selectors for the local polynomial CDF, PDF, and higher-order derivatives estimators implemented in \code{lpdensity()}. The selectors are rate-optimal for both interior and boundary evaluation points. Under an additional condition on the local polynomial fit discussed below, they are also consistent and hence (I)MSE-optimal. Both rule-of-thumb and direct plug-in implementations are available.
\end{itemize}

In addition, the methods \code{coef()}, \code{confint()}, \code{plot()}, \code{print()}, \code{summary()} and \code{vcov()} are supported for objects returned by \code{lpdensity()}, and the methods \code{coef()}, \code{print()} and \code{summary()} are supported for objects returned by \code{lpbwdensity()}. In particular, the function \code{plot()}, building on the \pkg{ggplot2} package in \proglang{R}, can be used to plot the estimated CDF, PDF, or higher-order derivatives for graphical illustration. This function takes the output from \code{lpdensity()}, and plots both point estimates and confidence intervals/bands for a collection of grid points. 

There are several other packages and functions available for kernel-based density estimation in \proglang{R}. Table \ref{table:comparison} gives a summary of their functionalities. As shown in that table, the package \pkg{lpdensity} is the first to offer consistent estimation of the CDF, PDF and density derivatives for both interior and boundary points, higher-order bias reduction, and valid inference both pointwise (confidence intervals) and uniformly (confidence bands). Section \ref{sec:simulation} compares the numerical performance of these packages in a simulation study.

\begin{table}[!t]
{\centering
\renewcommand{\arraystretch}{1.2}
\resizebox{\textwidth}{!}{
\begin{tabular}{lcccccc}
\hline\hline
Package                		&Density	 &Valid for   &Higher-order  &Standard      &Valid 		 & Confidence	\\
Function 	       			&derivative  &boundary    &bias reduction&error 	    &inference   & bands		\\ \hline\hline
\pkg{KernSmooth}    		&            &            &              &              &            &           	\\
~~\code{bkde}    			&$\times$    &$\times$    &$\times$      &$\times$      &$\times$    & $\times$		\\
~~\code{locpoly} 		    &$\checkmark$&$\checkmark$&$\checkmark$  &$\times$      &$\times$    & $\times$		\\ \hline
\pkg{ks} 					&            &            &              &              &            &				\\
~~\code{kdde} 				&$\checkmark$&$\times$    &$\times$      &$\times$      &$\times$    & $\times$		\\ 
~~\code{kde} 			    &$\times$    &$\times$    &$\times$      &$\times$      &$\times$    & $\times$		\\  \hline
\pkg{np}  					&            &            &              &              &            & 				\\
~~\code{npudens} 			&$\times$    &$\times$    &$\times$      & $\checkmark$ &$-$		 & $\times$		\\ 
~~\code{npuniden.boundary} 	&$\times$    &$\checkmark$&$\times$      &$\checkmark$  &$-$		 & $\times$		\\ \hline	
\pkg{nprobust}  			&            &            &              &              &            & 				\\
~~\code{kdrobust} 			&$\times$    &$\times$    &$\times$      &$\checkmark$  &$\checkmark$& $\times$		\\ \hline
\pkg{plugdensity} 			&            &            &              &              &            & 				\\
~~\code{plugin.density} 	&$\times$    &$\times$    &$\times$      &$\times$      &$\times$    & $\times$		\\ \hline
\code{stats::density} 		&$\times$    &$\times$    &$\times$      & $\times$     &$\times$    & $\times$		\\ \hline\hline
\pkg{lpdensity}          	&            &            &              &              &            & 				\\
~~\code{lpdensity}        	&$\checkmark$&$\checkmark$&$\checkmark$  &$\checkmark$  &$\checkmark$& $\checkmark$	\\
\hline\hline
\end{tabular}
}}
\caption{Comparison of \proglang{R} packages and functions.\label{table:comparison}}
\begin{flushleft}
\footnotesize{Notes. $\checkmark$ indicates the feature is available, $\times$ indicates the feature is not available, and $-$ indicates that inference is available and valid if undersmoothing is used but that is not the default in the package (and hence inference is invalid by default).}
\end{flushleft}
\end{table}

This article continues as follows. Section \ref{sec:overview} provides a brief, self-contained overview of the main ideas underlying the local polynomial estimators implemented in the package \pkg{lpdensity}. Section \ref{sec:illustration} illustrates the main features of our package. Section \ref{sec:simulation} showcases its finite-sample performance and compares it with other \proglang{R} packages implementing kernel-based density estimators. Section \ref{sec:conclusion} concludes. We also include two appendices. Appendix \ref{appendix:details} discusses in more detail our data-driven bandwidth selectors, and Appendix \ref{appendix:stata} illustrates the \proglang{Stata} implementation of the \pkg{lpdensity} package. Installation details, scripts replicating the numerical results reported herein, links to software repositories, and other companion information, can be found in the package's website: \url{https://nppackages.github.io/lpdensity/}.

%%%%%%%%%%%%%%%%%%%%%%%%%%%%%%%%%%%%%%%%%%%%%%%%%
%% SECTION: Overview                           %%
%%%%%%%%%%%%%%%%%%%%%%%%%%%%%%%%%%%%%%%%%%%%%%%%%
\section{Methodology and implementation}\label{sec:overview}

This section offers a brief overview of the main methods implemented in the \proglang{R} and \proglang{Stata} package \pkg{lpdensity}. For formal results, including assumptions, proofs and any other technical details see \citet*[hereafter CJM]{Cattaneo-Jansson-Ma_2020_JASA,Cattaneo-Jansson-Ma_2021_LocRegDistribution}.

We assume that $X_1,X_2,\cdots,X_n$ is a random sample from the random variable $X\in\mathcal{X}$, where $F(x)$ denotes its smooth CDF, $f(x)$ denotes its smooth PDF, and $\mathcal{X}\subseteq\mathbb{R}$ denotes its (possibly restricted) support, which can be bounded or unbounded. As it is well known, conventional kernel density estimators will be biased at or near boundary points, and other density estimators must be used if the goal is to estimate a density function on a compact support \citep[and references therein]{Karunamuni-Albert_2005_SM}. The package \pkg{lpdensity} implements a simple, easy-to-interpret and boundary adaptive density estimator based on local polynomial methods. As a by-product, the package also offers a smooth local polynomial estimate of the CDF as well as density derivatives. To cover all cases in an unified way, we employ the notation $g^{(\nu)}(x)=\partial^{\nu}g(x)/\partial x^{\nu}$ and $g(x)=g^{(0)}(x)$ for any smooth function $g(\cdot)$, and define $F(x)=F^{(0)}(x)$, $f(x)=F^{(1)}(x)$, and derivatives of the density function as $f^{(\nu-1)}(x)=F^{(\nu)}(x)$ with $\nu=1,2,\cdots$, with $f(x)=f^{(0)}(x)$. 

\subsection{Local polynomial distribution and density estimation}

To describe the estimators implemented in the package \pkg{lpdensity}, consider first the weighted empirical distribution function
\[
\hat{F}(x)=\frac{1}{n}\sum_{j=1}^{n}W_j\I(X_{j}\leq x),
\]
where $\I(\cdot)$ is the indicator function, and the weights $W_j$ are introduced for empirical applications such as missing data or counterfactual comparison (see CJM for examples). We assume these weights are normalized so that $\sum_{j=1}^n W_j/n=1$. The package \pkg{lpdensity} allows for a possibly estimated weighting scheme embedded in $\hat{F}(x)$, although for simplicity we will assume that each $W_i=1$ throughout this article. That is, $\hat{F}(x)$ is taken to be the standard root-$n$ consistent empirical distribution function estimator of $F(x)$.

As an alternative to conventional kernel density estimators, consider an estimator that first smooths out $\hat{F}(x)$ using local polynomials, and then constructs an estimator of $f(x)$ (and its derivatives). For $x\in\mathcal{X}$, the estimator implemented in the \pkg{lpdensity} package is
\begin{align*}
  \nonumber\widehat{\Bbeta}_{p}(x)
	&= \argmin_{\Bb\in \mathbb{R}^{p+1}}\sum_{i=1}^{n}
	  \left(\hat{F}(X_i)-\Br_{p}(X_i-x)^\Trans\Bb\right)^2 K\left(\frac{X_i-x}{h}\right) \\ 
	&= \begin{bmatrix}
	\frac{1}{0!} \hat{F}_p(x) & \frac{1}{1!} \hat{f}_p(x) & \frac{1}{2!}\hat{f}_p^{(1)}(x) & \cdots  & \frac{1}{p!} \hat{f}_p^{(p-1)}(x)
	\end{bmatrix}^\Trans, 
\end{align*}
where $\mathbf{r}_{p}(u)=(1,u,u^{2},\cdots ,u^{p})^\Trans$ is the $p$-th order polynomial expansion, $K(\cdot)$ is a kernel function such as the uniform or triangular kernel, and $h$ is a positive vanishing bandwidth sequence. The estimator approximates the discontinuous empirical CDF $\hat{F}(x)$ by a smooth local polynomial expansion using the weighting scheme implied by the kernel function, localized around the evaluation point $x$ according to the bandwidth $h$. CJM showed that
\[\widehat{\Bbeta}_{p}(x) \overset{\P}{\to} \Bbeta_p(x)
  = \begin{bmatrix}
  \frac{1}{0!} F(x) & \frac{1}{1!} f(x) & \frac{1}{2!} f^{(1)}(x) & \cdots & \frac{1}{p!} f^{(p-1)}(x)
  \end{bmatrix}^\Trans,
\]
as $h\to0$ and $nh^{2p-1}\to\infty$, where $\overset{\P}{\to}$ denotes convergence in probability. This implies that the least squares coefficients $\widehat{\Bbeta}_{p}(x)$ are consistent estimators of the CDF, PDF, and derivatives thereof at the evaluation point $x$.

Therefore, the generic local polynomial distribution estimator takes the form:
\[\hat{F}_{p}^{(\nu)}(x) = \hat{f}_{p}^{(\nu-1)}(x) = \nu! \Be_{\nu}^\Trans\widehat{\Bbeta}_{p}(x), \qquad 0\leq\nu\leq p,
\]
where $\mathbf{e}_{\nu}$ denotes the conformable unit vector that extracts the $(\nu +1)$-th estimated coefficient. This estimator is implemented in the function \code{lpdensity()}, given a choice of evaluation point $x$, polynomial degree $p$, derivative order $\nu$, kernel function $K$, and bandwidth $h$. In particular, the local polynomial density estimator is $\hat{f}_{p}(x)=\hat{F}_{p}^{(1)}(x)=\Be_1^\Trans\widehat{\Bbeta}_{p}(x)$, which is implemented via the default \code{lpdensity(...,p=2,v=1)}, employing a quadratic approximation to the empirical distribution function to construct the density estimator $\hat{f}_{2}(x)$. Similarly, higher-order derivatives of the CDF can be estimated through $\hat{f}^{(\nu-1)}_{p}(x)=\hat{F}^{(\nu)}_{p}(x)=\nu! \Be_{\nu}^\Trans\widehat{\Bbeta}_{p}(x)$ for $2\leq\nu\leq p$. We recommend using a local polynomial that is one order higher than the derivative to be estimated, $p=\nu+1$, or more generally to set $p-\nu$ odd. Of course, it is possible to achieve more bias reduction by increasing the local polynomial order $p$. 

The \code{lpdensity()} function employs the triangular kernel by default. Other available options include the uniform kernel and the Epanechnikov kernel. Generally speaking, the uniform kernel delivers a smaller asymptotic variance but a larger asymptotic bias for the resulting point estimator. It is possible reduce its asymptotic bias by using a kernel that is more concentrated around the origin, such as the triangular kernel, at the cost of increasing the asymptotic variance. The choice of the kernel, however, does not affect the orders of the bias and the variance, and hence this is less of a concern compared to bandwidth selection, which we discuss in more detail below. In addition, \cite*{Cattaneo-Jansson-Ma_2021_LocRegDistribution} show that more efficiency gains can be achieved by first including a higher-order polynomial term and then partialling out this term using a minimum distance approach. 

The rest of this section outlines the main properties, both statistical and numerical, of the estimator $\hat{F}^{(\nu)}_{p}(x)$, and discusses other related issues such as bandwidth selection, variance estimation, and valid (robust bias-corrected) inference. While the smooth CDF estimator $\hat{F}_{p}(x)=\hat{F}^{(0)}_{p}(x)$ is useful, the density and higher-order derivatives estimators are perhaps more relevant for empirical work. In particular, as mentioned before, the density estimator is intuitive and very easy to implement, while also being boundary adaptive. Thus, the estimator $\hat{f}_{p}(x)$ can be computed for all evaluation points on the (possibly restricted) support of $X$ in an automatic and straightforward way. This explains why the main functions in the package \pkg{lpdensity} refer to density estimation.

\subsection{Mean squared error}

The estimator $\hat{F}_{p}^{(\nu)}(x)$ can be written in the familiar weighted least squares form: $\widehat{\Bbeta}_p(x) = (\BX^\Trans\BW\BX)^{-1}(\BX^\Trans\BW\BY)$ with $\BX$ the usual polynomial design matrix and $\BW$ a diagonal matrix consisting of kernel weights. The only difference relative to standard local polynomial regression is that here the ``dependent variable'' is estimated: $\BY=[\hat{F}(X_1),\hat{F}(X_2),\cdots,\hat{F}(X_n)]^\Trans$, where $\hat{F}(x)$ is the possibly weighted empirical CDF. Unlike other local polynomial density estimators proposed in the literature \citep*[e.g.,][]{Cheng-Fan-Marron_1997_AoS,Zhang-Karunamuni_1998_JSPI}, the estimator $\hat{F}^{(\nu)}_{p}(x)$ does not require pre-binning or any other pre-processing of the data beyond constructing the empirical distribution function.  As a result, this estimation approach removes the need of choosing the number, position, and length of the bins in a preliminary histogram estimate. 

CJM obtained a general stochastic approximation to the bias and variance of the estimator $\hat{F}_{p}^{(\nu)}(x)$, $\nu=0,1,2,\cdots,p$, for all evaluation points $x\in\mathcal{X}$. Here we discuss the leading case of density estimation and derivatives thereof. For any choice of polynomial order $p$, and $1\leq\nu\leq p$, the variance and bias of $\hat{F}_p^{(\nu)}(x)$ are approximately
\begin{align*}
\mathsf{Var}[\hat{F}_p^{(\nu)}(x)] &= \frac{1}{nh^{2\nu-1}}\mathsf{V}_{\nu,p}(x), \\
\mathsf{Bias}[\hat{F}_p^{(\nu)}(x)] &= h^{p-\nu+1} \left[F^{(p+1)}(x)\mathsf{B}_{1,\nu,p}(x) + h \cdot F^{(p+2)}(x)\mathsf{B}_{2,\nu,p}(x)\right],
\end{align*}
where $\mathsf{V}_{\nu,p}(x)$, $\mathsf{B}_{1,\nu,p}(x)$, and $\mathsf{B}_{2,\nu,p}(x)$ denote quantities that can be constructed directly using only the data, choice of (preliminary) bandwidth, evaluation point, polynomial order, derivative order, and kernel function. That is, all these quantities are in pre-asymptotic form, which has been shown to offer higher-order distributional refinements in the context of local polynomial regression \citep*{Calonico-Cattaneo-Farrell_2018_JASA,Calonico-Cattaneo-Farrell_2020_CEOptimal}. Furthermore, it can be shown that $\mathsf{V}_{\nu,p}(x)$, $\mathsf{B}_{1,\nu,p}(x)$, and $\mathsf{B}_{2,\nu,p}(x)$ converge (in probability) to well-defined non-random limits. 

Since the above approximations are in pre-asymptotic form and are valid for all evaluation points, we can define a generic pointwise MSE-optimal bandwidth choice as
\[h_{\mathtt{MSE}, p}(x) = \argmin_{h>0} \mathsf{MSE}[\hat{F}_p^{(\nu)}(x)] = \argmin_{h>0} \left\{\mathsf{Var}[\hat{F}_p^{(\nu)}(x)] + \mathsf{Bias}[\hat{F}_p^{(\nu)}(x)]^2\right\}. 
\]
The optimal bandwidth also depends on $\nu$, but we suppress this in the notation to conserve notation. Under standard regularity conditions, $h_{\mathtt{MSE}, p}(x)$ is MSE optimal in rates for all evaluation points and choices of $p$ and $\nu$; and is MSE-optimal in constants if either (i) $p-\nu$ is odd or (ii) $x$ is a boundary point. See Appendix \ref{appendix:details} for more details.

We also define the IMSE-optimal bandwidth choice as follows:
\[h_{\mathtt{IMSE}, p} = \argmin_{h>0} \int\mathsf{MSE}[\hat{F}_p^{(\nu)}(x)]w(x)\mathrm{d}x, \]
where $w(x)$ denotes a user-chosen weighting scheme. Dependence on $\nu$ is again suppressed to ease notation. In practice, the integral will be approximated using the grid points specified in \code{lpdensity()} or \code{lpbwdensity()}, allowing for both a uniform weighting ($w(x)=1$) as well as the empirical distribution weighting.

The MSE-optimal and IMSE-optimal bandwidth selectors, $h_{\mathtt{MSE}, p}(x)$ and $h_{\mathtt{IMSE}, p}$, are carefully developed so that they automatically adapt to boundary points, while also retaining their main theoretical features (e.g., rate optimality). In practice, these bandwidths can be computed after replacing the unknown quantities by estimates thereof. We discuss implementation details below.

\subsection{Point estimation and robust bias-corrected inference}

Both $h_{\mathtt{MSE},p}(x)$ and $h_{\mathtt{IMSE},p}$, as well as their feasible counterparts, denoted by $\hat{h}_{\mathtt{MSE},p}(x)$ and $\hat{h}_{\mathtt{IMSE},p}$, can be used to construct MSE-optimal or IMSE-optimal point estimators for the PDF or its derivatives. The package \pkg{lpdensity} also allows for CDF estimation and computes an (I)MSE-optimal bandwidth, though we do not provide the details here to conserve space: some stochastic approximations change in non-trivial ways because the CDF estimator is $\sqrt{n}$-consistent. See CJM for more details.

As it is well known in the nonparametric literature, standard Wald-type inference is not valid when an (I)MSE-optimal bandwidth is used to construct the nonparametric point estimator. To be specific, the following distributional approximation holds for the standard Wald-type test statistic based on the local polynomial density estimator constructed using an (I)MSE-optimal bandwidth:
\[\mathsf{T}_{\nu,p}(x)
  = \frac{\hat{F}_p^{(\nu)}(x)-F^{(\nu)}(x)}
         {\sqrt{\mathsf{Var}[\hat{F}_p^{(\nu)}(x)]}}
    \rightsquigarrow \mathcal{N}(\mathsf{bias},\ 1), \qquad 1\leq\nu\leq p,
\]
where $\rightsquigarrow$ indicates convergence in distribution, and $\mathcal{N}(\mu, \sigma^2)$ denotes the normal distribution with mean $\mu$ and variance $\sigma^2$. The \textsf{bias} term cannot be dropped in general: if the point estimator $\hat{F}_p^{(\nu)}(x)$ is constructed using an (I)MSE-optimal bandwidth, inference based on the usual ``point estimator $\pm$ $z_{1-\alpha/2}$ $\times$ standard error'' confidence interval is invalid due to the presence of an asymptotic bias ($z_\alpha$ denotes the $\alpha$ quantile of the standard normal distribution). The mechanical solution to this inferential problem is to undersmooth the point estimator $\hat{F}_p^{(\nu)}(x)$ using an \textit{ad hoc} bandwidth $h$ smaller than $h_\mathtt{MSE}(x)$ or $h_\mathtt{IMSE}$. Of course, the function \code{lpdensity()} allows for this approach by simply running bandwidth selection, estimation, and inference in separate steps (see Section \ref{sec:illustration} for an illustration).

\citet*{Calonico-Cattaneo-Farrell_2018_JASA,Calonico-Cattaneo-Farrell_2020_CEOptimal} showed that undersmoothing is suboptimal for inference under the same assumptions employed to construct an (I)MSE-optimal bandwidth. Instead, it is demonstrably better, in terms of higher-order distributional approximations and asymptotic coverage properties, to employ robust bias correction (RBC). The idea is to bias correct the point estimator and then adjust the variance accordingly. Heuristically, and abusing notation for simplicity, this leads to the Wald-type test statistic
\[\mathsf{T}^{\mathtt{RBC}}_{\nu,p}(x)
  = \frac{\hat{F}^{(\nu), \mathtt{BC}}_p(x)-F^{(\nu)}(x)}
         {\sqrt{\mathsf{Var}[\hat{F}^{(\nu), \mathtt{BC}}_p(x)]}}
    \rightsquigarrow \mathcal{N}(0,1),\qquad \hat{F}^{(\nu), \mathtt{BC}}_p(x) = \hat{F}^{(\nu)}_p(x) - \mathsf{Bias}[\hat{F}_p^{(\nu)}(x)],
\]
which has a valid standard normal distribution even when an MSE, IMSE or a cross-validation-type bandwidth for $\hat{F}_p^{(\nu)}(x)$ is used. Confidence intervals with correct asymptotic coverage can be constructed by inverting the test statistic $\mathsf{T}^{\mathtt{RBC}}_{\nu,p}(x)$. In particular, it can be shown that a RBC confidence interval is equivalent to employing the test statistic $\mathsf{T}_{\nu,p+1}(x)=\mathsf{T}^{\mathtt{RBC}}_{\nu,p}(x)$ for a particular choice of parameters/implementation. Therefore, the function \code{lpdensity()} employs an RBC test statistic by default, assuming an (I)MSE-optimal or cross-validation-based bandwidth for the $p$-th order point estimator is used, denoted generically by $h_{p}$, and therefore forms the test statistic
\[\mathsf{T}^{\mathtt{RBC}}_{\nu,p}(x)\equiv \mathsf{T}_{\nu,p+1}(x;h_p)
  = \frac{\hat{F}^{(\nu)}_{p+1}(x;h_p)-F^{(\nu)}(x)}
         {\sqrt{\mathsf{Var}[\hat{F}^{(\nu)}_{p+1}(x;h_p)]}},
\]
and associated confidence intervals
\[\mathsf{CI}_{\nu,p}^{\mathtt{RBC}}(x)\equiv \mathsf{CI}_{\nu,p+1}(x;h_p)
  = \left[\hat{F}^{(\nu)}_{p+1}(x;h_p)\ \  \pm\ \  z_{1-\alpha/2} \sqrt{\mathsf{Var}[\hat{F}^{(\nu)}_{p+1}(x;h_p)]}\right].
\]
The notation makes the bandwidth explicit to distinguish the two polynomial degrees used in constructing the point estimator and the RBC confidence interval/test statistic: (i) a $p$-th order polynomial is used for point estimation (and bandwidth selection), and (ii) a $(p+1)$-th order polynomial is used for inference. 

More generally, the package \pkg{lpdensity} implements confidence intervals of the form:
\[\mathsf{CI}_{\nu,p}^{\mathtt{RBC},q}(x)\equiv \mathsf{CI}_{\nu,q}(x;h_p)
  = \left[\hat{F}^{(\nu)}_{q}(x;h_p)\ \  \pm\ \  z_{1-\alpha/2} \sqrt{\mathsf{Var}[\hat{F}^{(\nu)}_{q}(x;h_p)]}\right],
\]
with $q$ determining the inference approach. The above confidence interval is thus based on  inverting the statistic $\mathsf{T}^{\mathtt{RBC},q}_{\nu,p}(x)\equiv \mathsf{T}_{\nu,q}(x;h_p)$, and by default we set $q=p+1$. CJM formally showed that the RBC confidence intervals have asymptotically correct coverage:
\[ \lim_{n\to\infty}\mathbb{P}\left[ F^{(\nu)}(x) \in \mathsf{CI}_{\nu,p}^{\mathtt{RBC},q}(x) \right]= 1-\alpha,\qquad \forall x\in\mathcal{X}. \]

In addition to pointwise confidence intervals, the \pkg{lpdensity} package also offers uniform confidence bands for the CDF, PDF, or derivatives thereof. The uniform confidence band for $F^{(\nu)}(x)$ takes a similar form,
\[\mathsf{CB}_{\nu,p}^{\mathtt{RBC},q}(\mathcal{G})
  \equiv \mathsf{CB}_{\nu,q}(\mathcal{G};h_p)
  = \left\{\left[\hat{F}^{(\nu)}_{q}(x;h_p)\ \  \pm\ \  z_{\mathcal{G},1-\alpha/2} \sqrt{\mathsf{Var}[\hat{F}^{(\nu)}_{q}(x;h_p)]}\right],\quad x\in\mathcal{G}\right\},
\]
with two noticeable differences. First, the confidence band no longer depends on the evaluation point, but rather on a collection of evaluation points, $\mathcal{G}$. Second, the critical value also changes, which is now denoted by $z_{\mathcal{G},1-\alpha/2}$. In practice, the new critical value can be obtained by first simulating a suitable Brownian bridge on the grid $\mathcal{G}$, and then computing the upper $\alpha$ quantile of the supremum of the simulated process. The option \code{CIuniform=TRUE} enables estimation and reporting of the uniform confidence band, which is turned off by default. CJM established a uniformly valid distributional approximation for the stochastic process $\{\mathsf{T}^{\mathtt{RBC},q}_{\nu,p}(x):x\in\mathcal{G}\}$, and proved that a nominal $1-\alpha$ level RBC confidence band is asymptotically valid:
\[ \lim_{n\to\infty}\mathbb{P}\left[ F^{(\nu)}(x) \in \mathsf{CI}_{\nu,p}^{\mathtt{RBC},q}(\mathcal{G}),\ \forall x\in\mathcal{G} \right] = 1-\alpha. \]
See \citet*{Cattaneo-Jansson-Ma_2021_LocRegDistribution} for technical details, regularity conditions, and additional discussions.  

Robust bias correction methods lead to confidence intervals/bands that will not be centered at the density point estimates because of the recentering introduced by the bias correction. That is, different polynomial orders are used for constructing point estimates and confidence intervals/bands. Setting $q$ and $p$ to be equal delivers confidence intervals/bands that are centered at the point estimates, but requires undersmoothing for valid inference (i.e., an (I)MSE-optimal bandwdith cannot be used). Hence the bandwidth would need to be specified manually when $q=p$, and the point estimates will no longer be (I)MSE-optimal. Sometimes the point estimates may even lie outside of the confidence intervals/bands, which can happen if the underlying distribution exhibits high curvature at some evaluation point(s). One possible solution in this case is to increase the polynomial order $p$ or to employ a smaller bandwidth.

\subsection{Bandwidth selection}

The package \pkg{lpdensity} implements several bandwidth selectors through \code{lpbwdensity()}, including MSE-optimal and IMSE-optimal plug-in rules, as well as rule-of-thumb bandwidth selectors based on a normal reference model. We only outline the main aspects of bandwidth selection here, but further details are given in Appendix \ref{appendix:details}.

To introduce our bandwidth selectors, recall that the quantities $\mathsf{V}_{\nu,p}(x)$, $\mathsf{B}_{1,\nu,p}(x)$, and $\mathsf{B}_{2,\nu,p}(x)$ are given in pre-asymptotic form, and hence they can be computed from the data directly given a pilot/preliminary bandwidth. As a consequence, to construct the (I)MSE-optimal bandwidth, the only unknown quantities are $F^{(p+1)}(x)$ and $F^{(p+2)}(x)$, which can be consistently estimated using the local polynomial density derivative estimators implemented in \code{lpdensity()} with a pilot/preliminary bandwidth. To be more precise, the MSE-optimal bandwidth is estimated by 
\[\hat{h}_{\mathtt{MSE}, p}(x)
  = \argmin_{h>0} \left\{\mathsf{Var}[\hat{F}_p^{(\nu)}(x)] + \widehat{\mathsf{Bias}}[\hat{F}_p^{(\nu)}(x)]^2\right\},
\]
with $\widehat{\mathsf{Bias}}[\hat{F}_p^{(\nu)}(x)]$ constructed by replacing $F^{(p+1)}(x)$ and $F^{(p+2)}(x)$ with their estimated counterparts. Similarly, the IMSE-optimal bandwidth selector is given by
\[\hat{h}_{\mathtt{IMSE}, p}
  = \argmin_{h>0} \sum_{g_j\in \mathcal{G}}\left\{\mathsf{Var}[\hat{F}_p^{(\nu)}(g_j)] + \widehat{\mathsf{Bias}}[\hat{F}_p^{(\nu)}(g_j)]^2\right\}, 
\]
where $\mathcal{G}$ is the collection of grid points specified in the function (by default, $\mathcal{G}$ takes on nineteen quantile-spaced values over the support of the data). 

\subsection{CDF estimation}

While the estimator $\hat{F}_p^{(\nu)}(x)$ is valid for all $\nu\geq0$, our discussion so far focused on the case $\nu\geq1$ because the resulting estimators of the density ($\nu=1$) and its derivatives ($\nu\geq2$) are the main focus of the package. Nevertheless, as a by-product, CJM developed analogous estimation, bandwidth selection and RBC inference results for the smooth CDF estimator $\hat{F}_p(x)=\hat{F}_p^{(0)}(x)$. These results are also implemented in the package \pkg{lpdensity} via the option \code{v=0}. For example, CDF estimation using a local constant approximation is obtained using \code{lpdensity(...,p=0,v=0)}, which employs the corresponding MSE-optimal bandwidth (\code{bwselect="mse-dpi"}) and a local linear approximation for inference (\code{q=p+1}) by default.

%%%%%%%%%%%%%%%%%%%%%%%%%%%%%%%%%%%%%%%%%%%%%%%%%
%% SECTION: Numerical                          %%
%%%%%%%%%%%%%%%%%%%%%%%%%%%%%%%%%%%%%%%%%%%%%%%%%
\section{Implementation and numerical illustration}\label{sec:illustration}

\begin{figure}[t!]
\centering
\centering
\subfloat[Histogram.]{\resizebox{0.45\columnwidth}{!}{\includegraphics{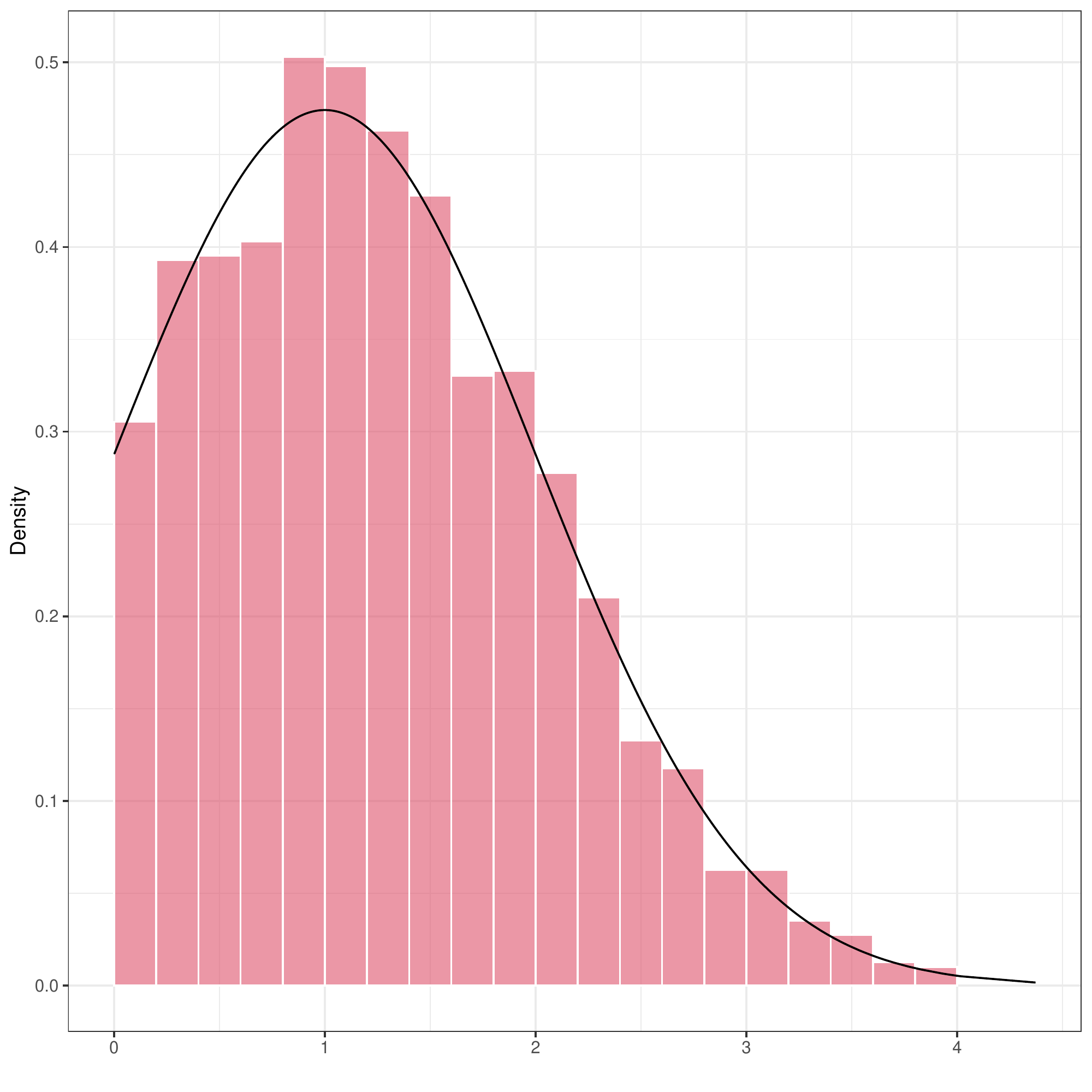}}}
\subfloat[Density plot.]{\resizebox{0.46\columnwidth}{!}{\includegraphics{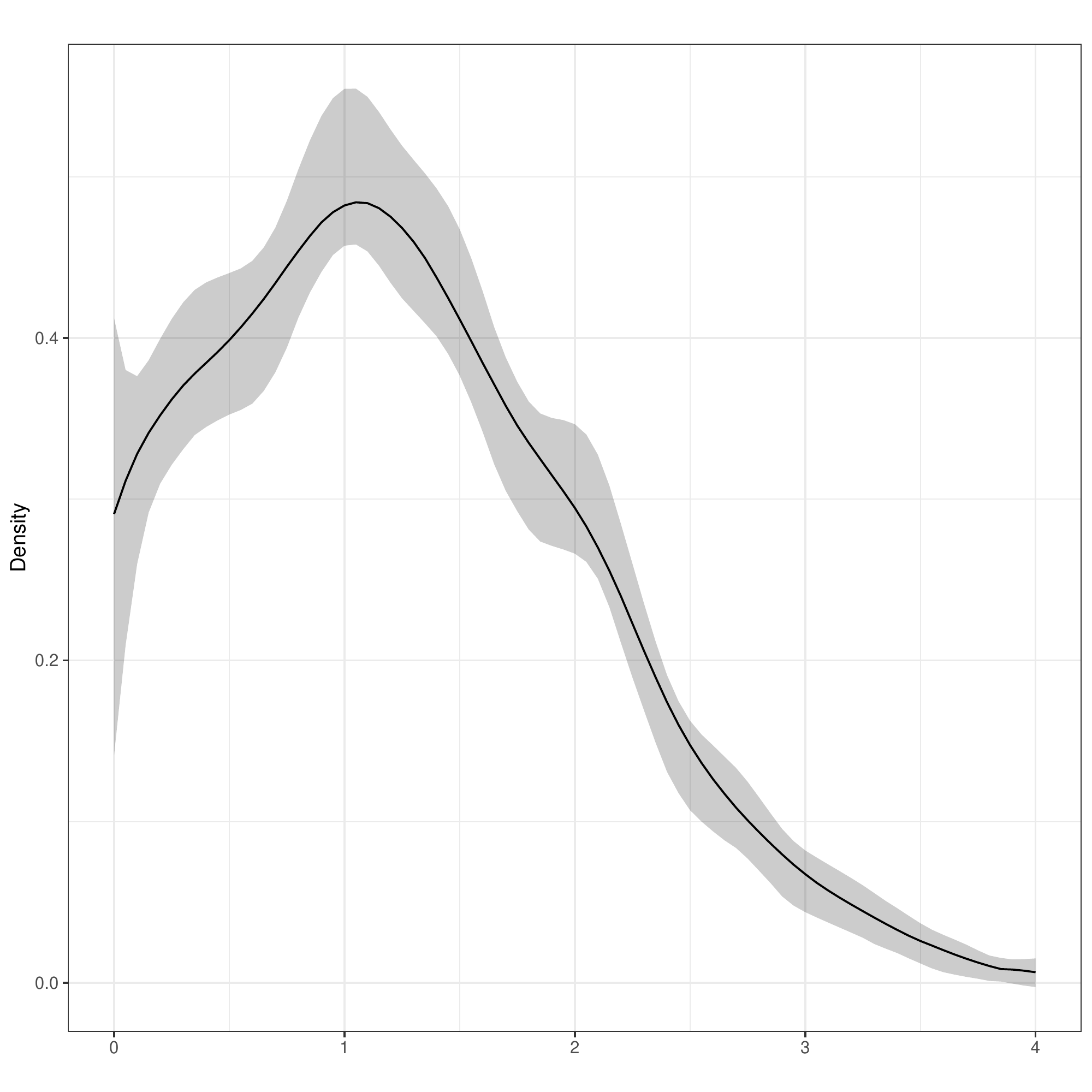}}}
\caption{Histogram of the simulated data and the density plot.}\label{fig: 1}
\end{figure}

We showcase some of the main features of the \pkg{lpdensity} package. The data consists of 2,000 observations simulated from the normal distribution $\mathcal{N}(1,1)$ truncated from below at 0. We create a discontinuity in density at $x=0$ to illustrate the performance of our procedure at boundaries. Panel (a) of Figure \ref{fig: 1} plots a histogram estimate and the true density function.
\begin{CodeChunk}
\begin{CodeInput}
R> set.seed(42)
R> data <- rnorm(4000, mean = -1)
R> data <- data[data < 0]; data <- -1 * data[1:2000]
\end{CodeInput}
\end{CodeChunk}

\subsection[Function lpdensity()]{Function \code{lpdensity()}}

The function \code{lpdensity()} provides both point estimates as well as RBC inference (confidence intervals and bands) employing the local polynomial density estimator, given a grid of points and a bandwidth choice. If the latter are not provided, then by default the function chooses nineteen quantile-spaced grid points over the support of the data and computes $\hat{h}_{\mathtt{MSE},p}(x)$ at each point.

The following command estimates the density function (\code{v=1}, the default) with fixed bandwidth \code{bw=0.5} at points $0, 0.5, \cdots, 4$, using a local quadratic approximation (\code{p=2}, the default) to the empirical distribution function. RBC confidence intervals over the grid are also computed, in this case using a local cubic approximation (\code{q=3}, the default). 
\begin{CodeChunk}
\begin{CodeInput}
R> model1 <- lpdensity(data, bw = 0.5, grid = seq(0, 4, 0.5))
R> summary(model1)
\end{CodeInput}
\begin{CodeOutput}
Call: lpdensity

Sample size                              (n=)    2000
Polynomial order for point estimation    (p=)    2
Density function estimated               (v=)    1
Polynomial order for confidence interval (q=)    3
Kernel function                                  triangular
Bandwidth selection method                       user provided

===========================================================================
                                     Point      Std.       Robust B.C.       
Index     Grid      B.W.   Eff.n      Est.     Error      [ 95% C.I. ]       
===========================================================================
1       0.0000    0.5000     355    0.2908    0.0436     0.1413 ,  0.4121    
2       0.5000    0.5000     799    0.3986    0.0147     0.3525 ,  0.4402    
3       1.0000    0.5000     919    0.4822    0.0160     0.4572 ,  0.5545    
4       1.5000    0.5000     820    0.4116    0.0150     0.3767 ,  0.4675    
5       2.0000    0.5000     564    0.2946    0.0137     0.2662 ,  0.3465    
---------------------------------------------------------------------------
6       2.5000    0.5000     320    0.1475    0.0099     0.1071 ,  0.1626    
7       3.0000    0.5000     147    0.0674    0.0069     0.0438 ,  0.0821    
8       3.5000    0.5000      59    0.0259    0.0045     0.0120 ,  0.0369    
9       4.0000    0.5000      15    0.0065    0.0022    -0.0027 ,  0.0151    
===========================================================================
\end{CodeOutput}
\end{CodeChunk}
The first part of the output provides basic information on the options specified in the function. For example, the default estimand is the density function, indicated by \code{Density function estimated (v=) 1}. The rest of the output gives estimation results, including (i) \code{Grid}: the grid points; (ii) \code{B.W.}: the bandwidths; (iii) \code{Eff.n}: the effective sample size for each grid point; (iv) \code{Point Est.}: the point estimates using polynomial order \code{p}, and the associated standard errors under \code{Std.} \code{Error}; (v) \code{Robust B.C.[95\% C.I.]}: robust bias-corrected 95\% confidence intervals. Point estimates, standard errors, and other information can be easily extracted for further statistical analysis. The output is stored in a standard matrix, and can be accessed with the following.
\begin{CodeChunk}
\begin{CodeInput}
R> model1$Estimate
\end{CodeInput}
\end{CodeChunk}
When the argument \code{grid} is suppressed, the evaluation points will be the $0.05, 0.1, \cdots,0.9, 0.95$ quantiles computed from the data. Conventional inference results (i.e., without robust bias correction) can be obtained by setting \code{q=p}. For example (output is suppressed):
\begin{CodeChunk}
\begin{CodeInput}
R> summary(lpdensity(data, bw = 0.5, p = 2, q = 2))
\end{CodeInput}
\end{CodeChunk}
It is also possible to suppress the argument \code{bw}, and the function will select the bandwidth automatically by minimizing (an estimated approximation to) the mean squared error, employing \code{lpbwdensity()}. Other bandwidth selection methods are available; we will illustrate data-driven bandwidth selection procedures in an upcoming subsection.

The method \code{summary()} takes six additional arguments. The first one, \code{alpha}, specifies the (one minus) nominal coverage of the confidence interval, with default being 0.05. Another argument is \code{sep}, which controls the horizontal separator. The default value is 5, and hence a dashed line is drawn after every five grid points. This feature can be suppressed by setting it to 0. Sometimes it may be desirable to report only a subset of the estimates, which can be done by using either the \code{grid} or the \code{gridIndex} option. The \code{grid} option allows reporting results for a selected set of grid points originally specified in the \code{lpdensity()} function, while \code{gridIndex} helps achieve the same goal by specifying the indices of the grid points. The last two options are related to confidence bands. By setting \code{CIuniform=TRUE}, a uniform confidence band, instead of pointwise confidence intervals, will be reported. Because the critical values have to be simulated in this case, the number of simulations used is controlled by the option \code{CIsimul} (its default value is 2000). The following example produces the 99\% confidence band for four grid points $0$, $0.5$, $1$ and $2$, with dashed lines appearing after every three grid points. (Fixing the random seed allows reproducing the simulated critical values and the confidence intervals.)
\begin{CodeChunk}
\begin{CodeInput}
R> set.seed(123)
R> summary(model1, alpha = 0.01, sep = 3, grid = c(0, 0.5, 1, 2), 
+    CIuniform = TRUE)
\end{CodeInput}
\begin{CodeOutput}
Call: lpdensity

Sample size                              (n=)    2000
Polynomial order for point estimation    (p=)    2
Density function estimated               (v=)    1
Polynomial order for confidence interval (q=)    3
Kernel function                                  triangular
Bandwidth selection method                       user provided

===========================================================================
                                     Point     Std.        Robust B.C.       
Index     Grid      B.W.   Eff.n      Est.     Error   [ Unif. 99% C.I. ]    
===========================================================================
1       0.0000    0.5000     355    0.2908    0.0436     0.0606 ,  0.4927    
2       0.5000    0.5000     799    0.3986    0.0147     0.3263 ,  0.4664    
3       1.0000    0.5000     919    0.4822    0.0160     0.4283 ,  0.5835    
---------------------------------------------------------------------------
5       2.0000    0.5000     564    0.2946    0.0137     0.2423 ,  0.3704    
===========================================================================
\end{CodeOutput}
\end{CodeChunk}

Another important argument in \code{lpdensity()} is \code{scale}, which scales the point estimates and standard errors. This is particularly useful if only part of the data is used. 
For example, assume one would like to estimate the PDF using the two subsamples $\{X_i:X_i<1.5\}$ and $\{X_i:X_i>1.5\}$ separately. Simply splitting the data will not give consistent estimates, as it produces conditional (rather than marginal) density estimates:
\begin{CodeChunk}
\begin{CodeInput}
R> lpdensity(data[data < 1.5], bw = 0.5, grid = 1.5)$Estimate[, "f_p"]
R> lpdensity(data[data > 1.5], bw = 0.5, grid = 1.5)$Estimate[, "f_p"]
R> dnorm(1.5, mean = 1, sd = 1) / pnorm(0, mean = 1, sd = 1, 
+    lower.tail = FALSE)
\end{CodeInput}
\begin{CodeOutput}
[1] 0.6755464
[1] 1.222052
[1] 0.4184555
\end{CodeOutput}
\end{CodeChunk}
The previous commands give point estimates $0.676$ and $1.222$, which are far from the true value $0.418$. To have consistent estimates, we need to scale the estimates by the proportion of the data used for estimation: 
\begin{CodeChunk}
\begin{CodeInput}
R> lpdensity(data[data < 1.5], bw = 0.5, grid = 1.5, 
+    scale = sum(data < 1.5)/2000)$Estimate[, "f_p"]
R> lpdensity(data[data > 1.5], bw = 0.5, grid = 1.5, 
+    scale = sum(data > 1.5)/2000)$Estimate[, "f_p"]
\end{CodeInput}
\begin{CodeOutput}
[1] 0.4303231
[1] 0.443605
\end{CodeOutput}
\end{CodeChunk}

\subsection[Function plot()]{Function \code{plot()}} 

The function \code{plot()}, along with many other methods, is supported. This function takes the output from \code{lpdensity()} and produces plots of point estimates and robust bias-corrected confidence intervals/bands over the grid of evaluation points selected. Panel (b) of Figure \ref{fig: 1} shows how plots can be easily generated.
\begin{CodeChunk}
\begin{CodeInput}
R> model2 <- lpdensity(data, bw = 0.5, grid = seq(0, 4, 0.05))
R> plot(model2) + theme(legend.position = "none")
\end{CodeInput}
\end{CodeChunk}
The confidence intervals/bands are not centered at the point estimates in general. As described in Section \ref{sec:overview}, by default the point estimates are constructed using MSE-optimal bandwidths, which implies the smoothing bias is non-negligible and hence valid inference should be based on robust bias-corrected confidence intervals. 

\begin{figure}[t!]
\centering
\subfloat[Dashed confidence intervals.]{\resizebox{0.45\columnwidth}{!}{\includegraphics{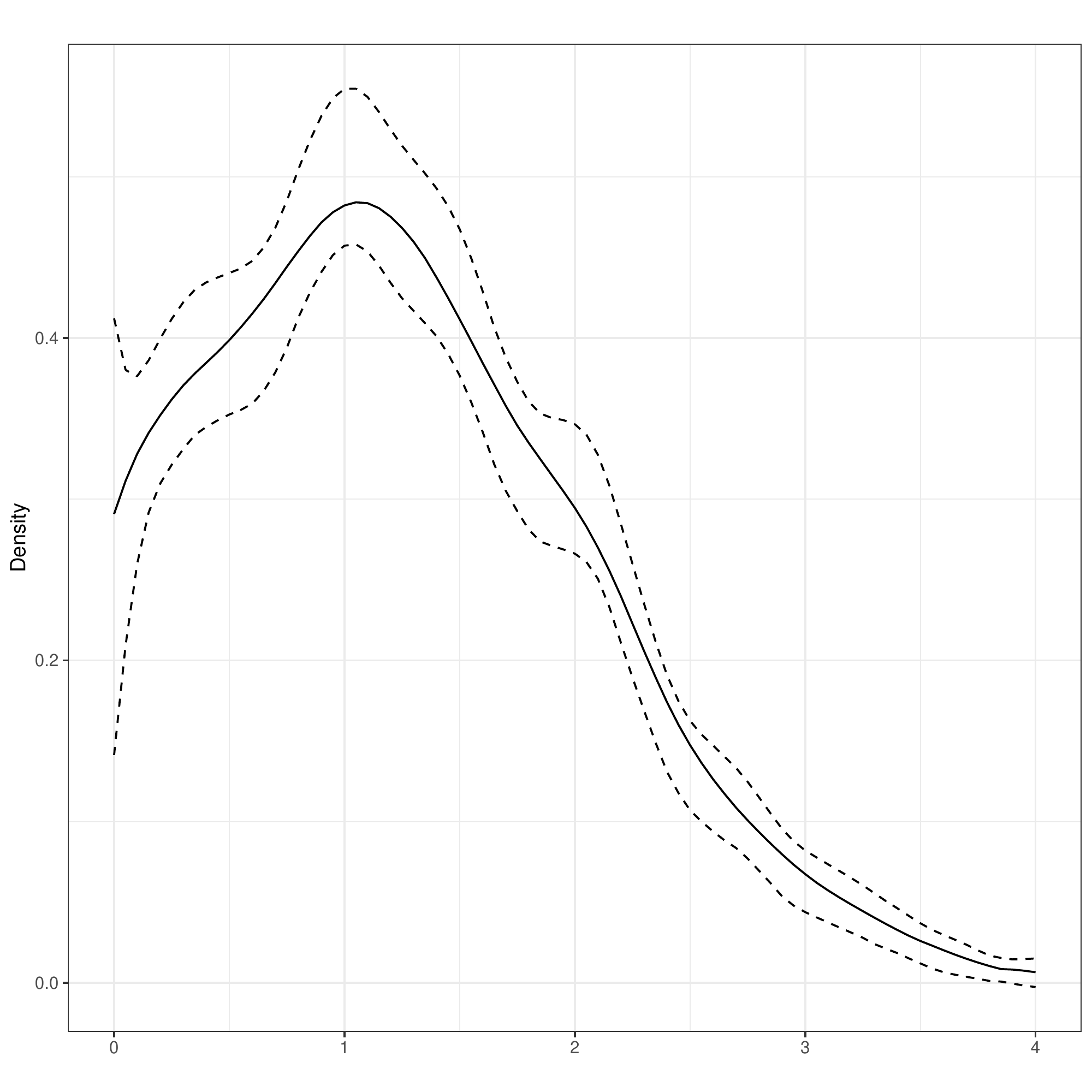}}}
\subfloat[Error bars for a subset of grid points.]{\resizebox{0.45\columnwidth}{!}{\includegraphics{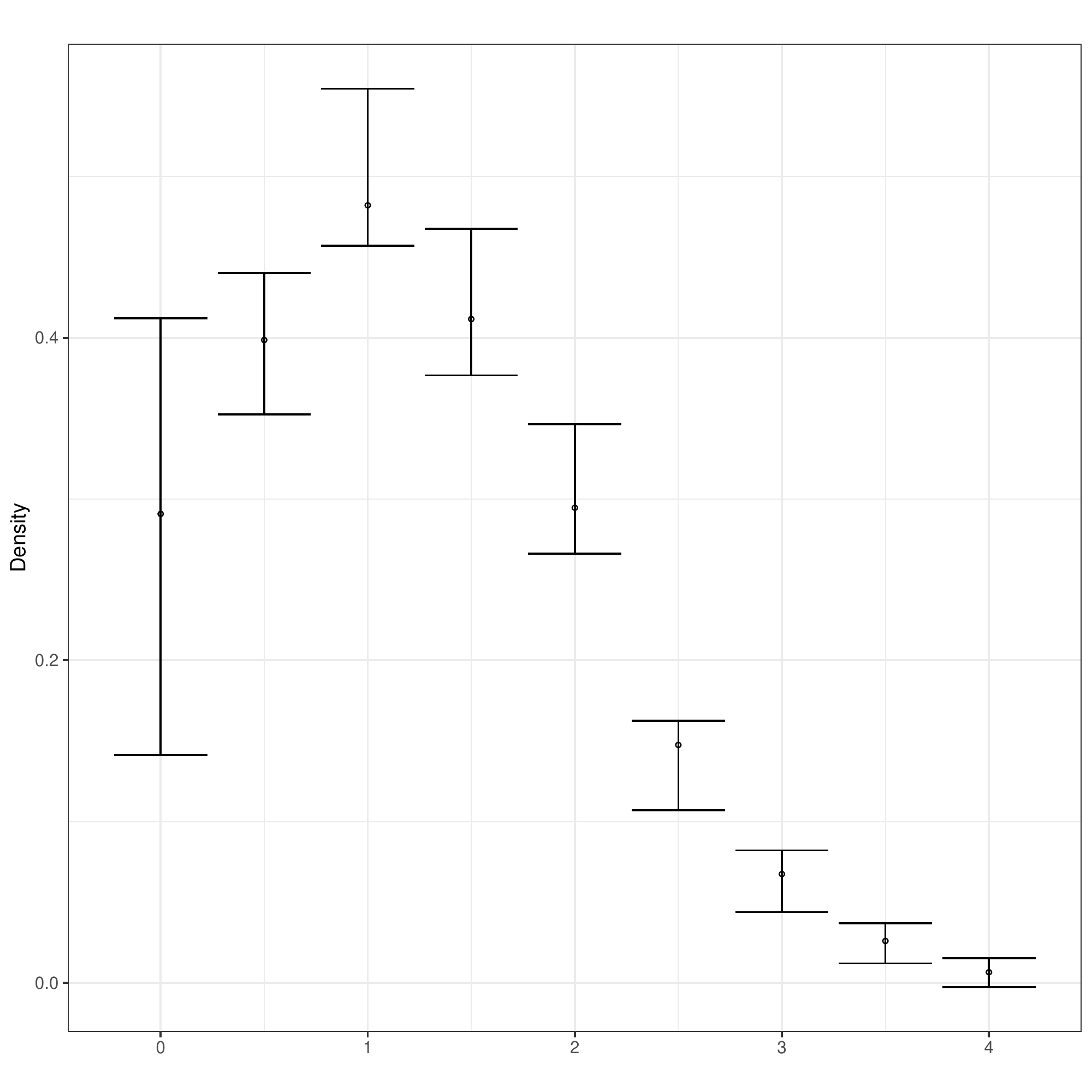}}}\\
\subfloat[Histogram.]{\resizebox{0.45\columnwidth}{!}{\includegraphics{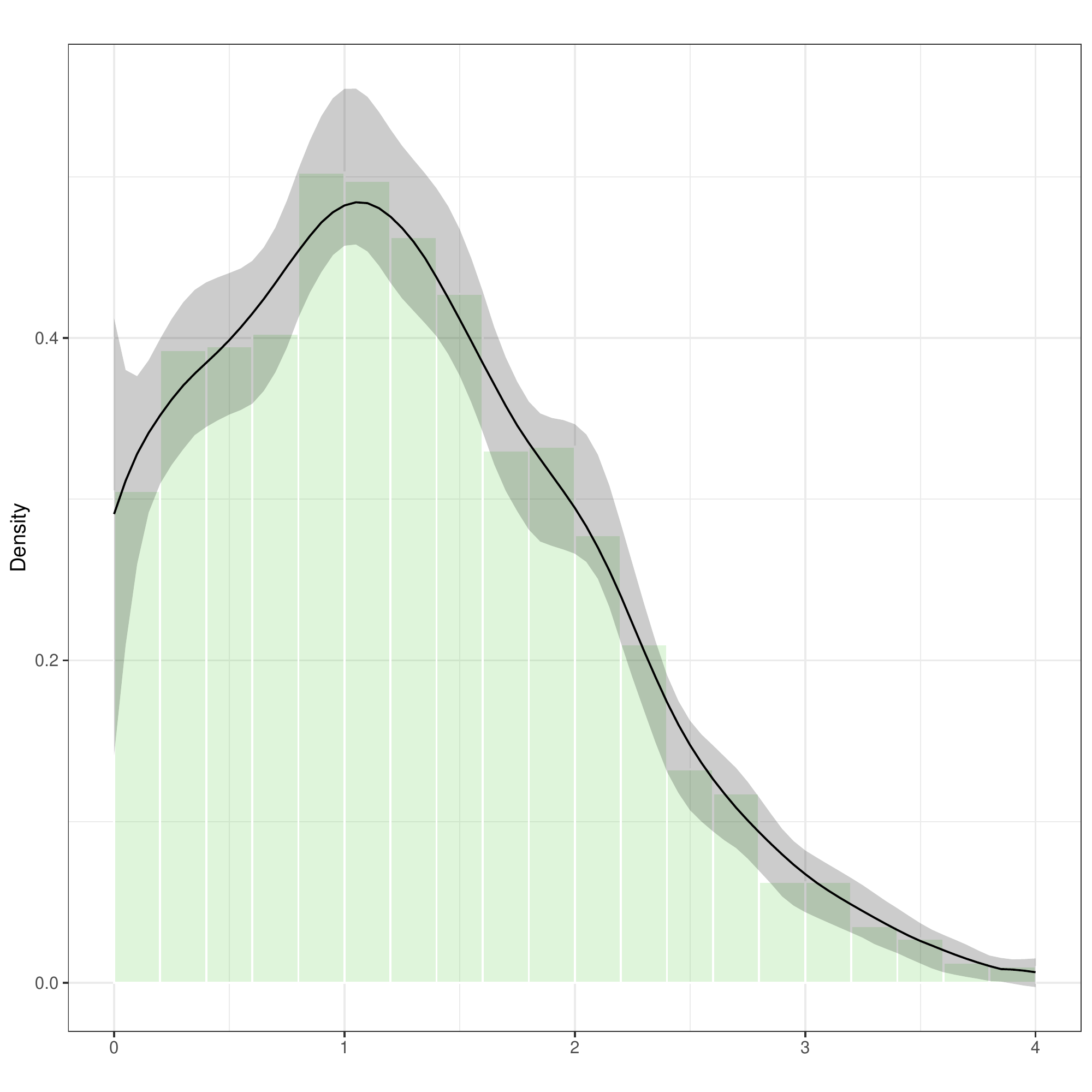}}}
\subfloat[90\% uniform confidence band.]{\resizebox{0.45\columnwidth}{!}{\includegraphics{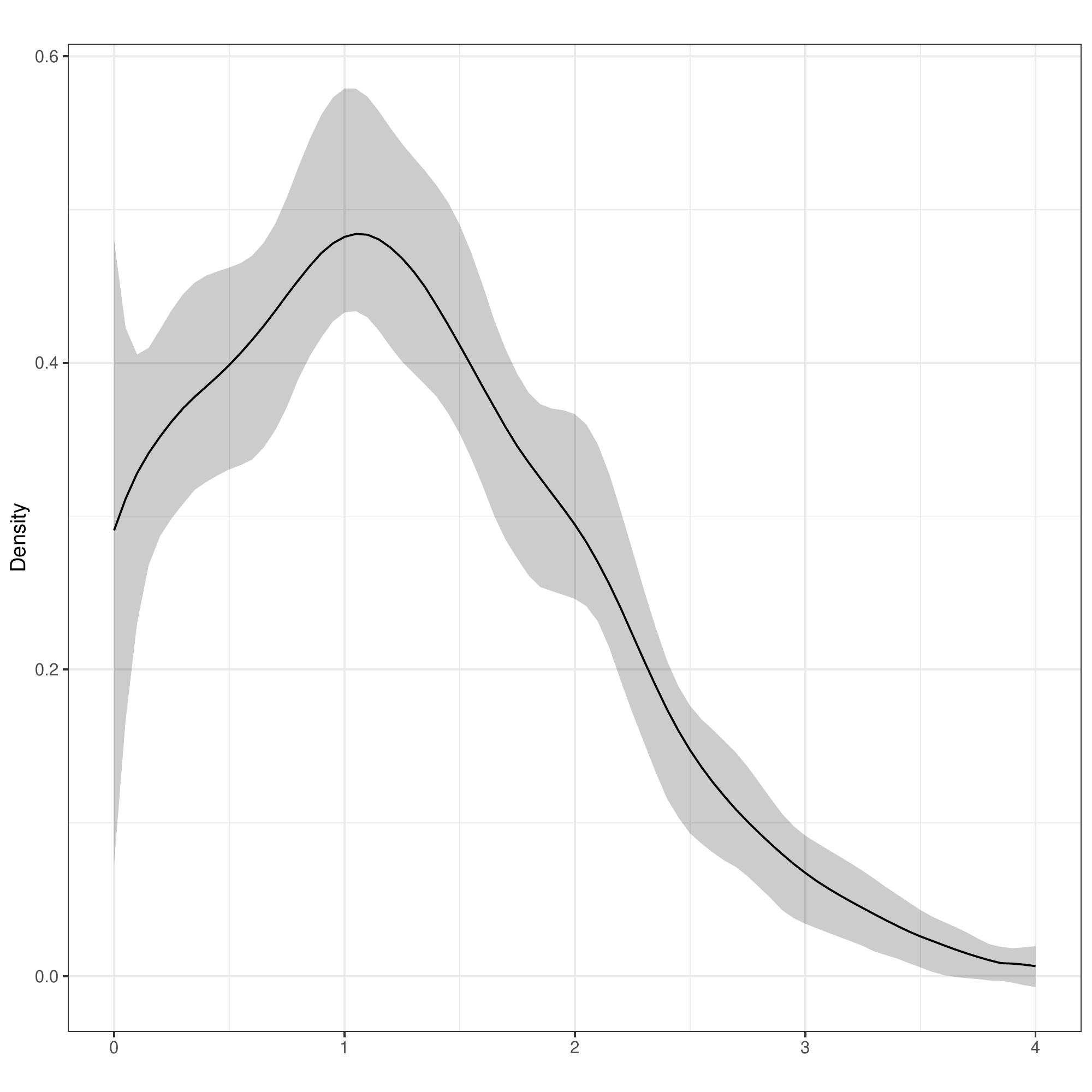}}}
\caption{Density plots with different specifications. }\label{fig: 2}
\end{figure}

The function \code{plot()} allows for customization: Figure \ref{fig: 2} illustrates some of the features. (For panel (d), we again fix the random seed to reproduce the simulated critical values and the confidence band.)
\begin{CodeChunk}
\begin{CodeInput}
R> plot(model2, CItype = "line") + theme(legend.position = "none")
R> plot(model2, type="points", CItype="ebar", grid = seq(0, 4, 0.5)) +
+    theme(legend.position = "none") 
R> plot(model2, hist = TRUE, histData = data, histBreaks = seq(0, 4, 0.2)) + 
+    theme(legend.position = "none")
R> set.seed(123)
R> lpdensity.plot(model2, alpha = 0.1, CIuniform = TRUE) + 
+    theme(legend.position = "none")
\end{CodeInput}
\end{CodeChunk}

\subsection[Function lpbwdensity()]{Function \code{lpbwdensity()}}

The function \code{lpbwdensity()} implements four bandwidth selectors, (i) MSE-optimal plug-in bandwidth selector, denoted by \code{"mse-dpi"} (this is the default option), (ii) IMSE-optimal plug-in bandwidth selector, denoted by \code{"imse-dpi"}, (iii) rule-of-thumb bandwidth selector with a normal reference model, denoted by \code{"mse-rot"}, and (iv) integrated rule-of-thumb bandwidth selector, denoted by \code{"imse-rot"}. We illustrate some of the main features of \code{lpbwdensity()} with the same simulated data used previously.

\begin{figure}[t!]
\centering
\subfloat[IMSE-bandwidth]{\resizebox{0.45\columnwidth}{!}{\includegraphics{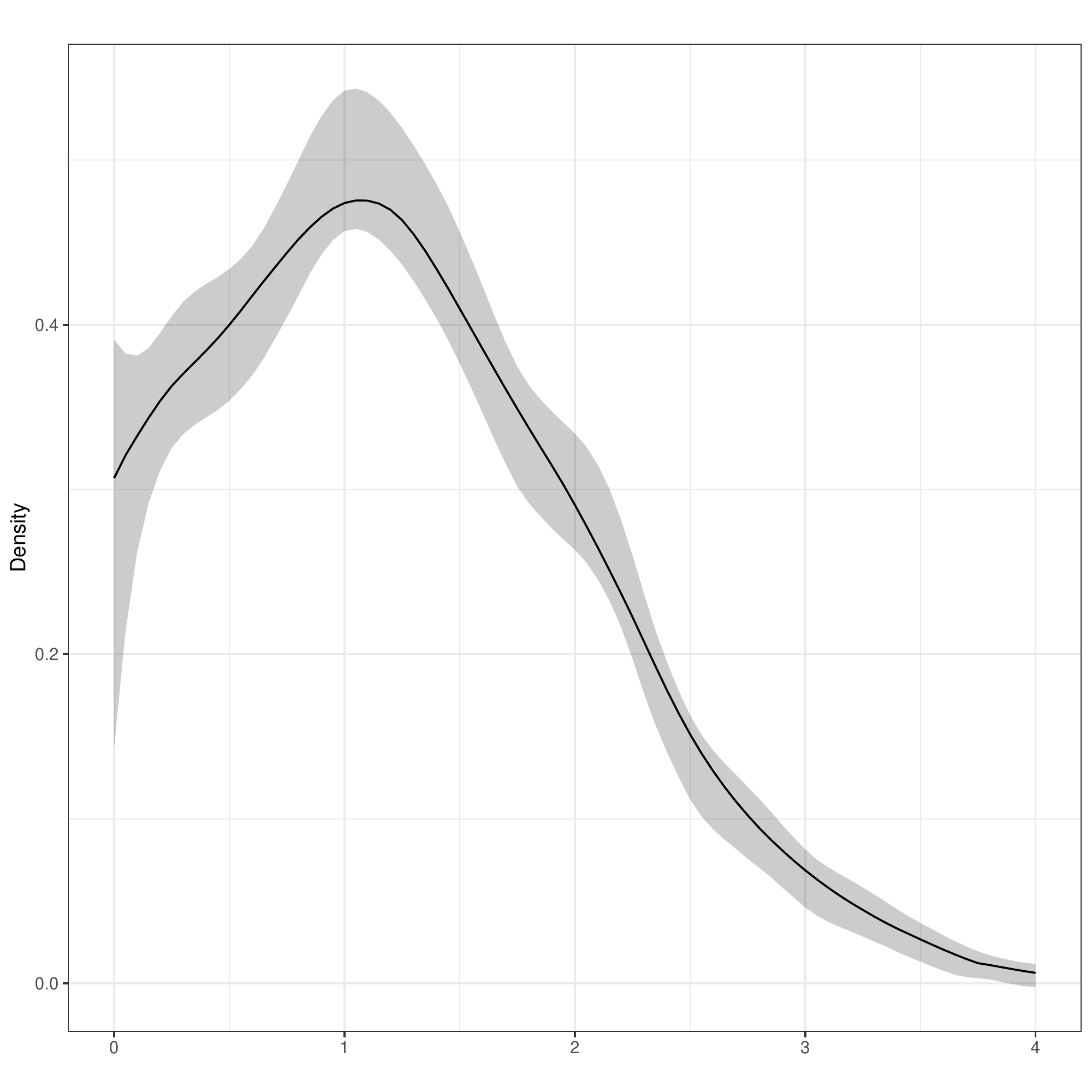}}}
\subfloat[Under-smoothing]{\resizebox{0.45\columnwidth}{!}{\includegraphics{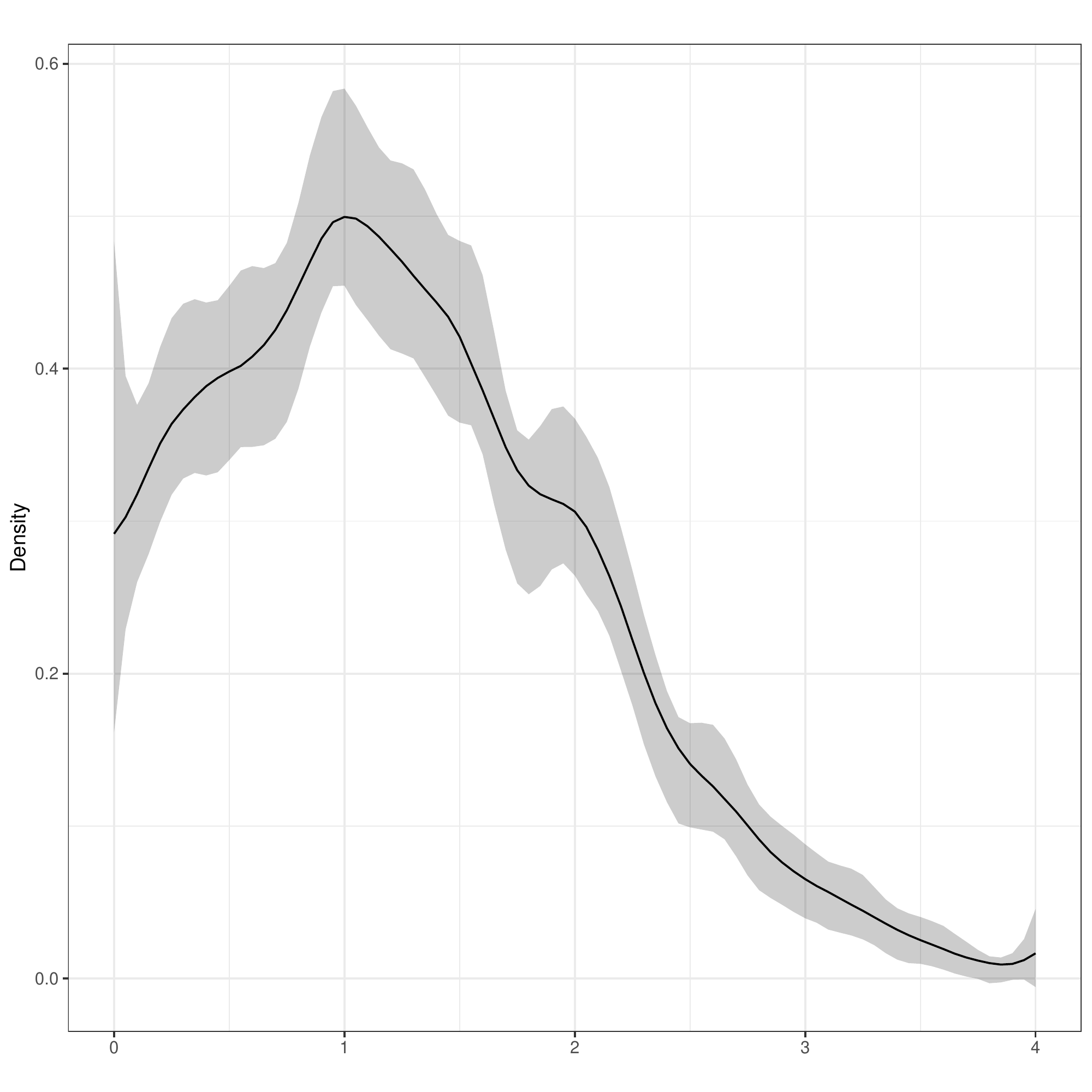}}}
\caption{Density plot with IMSE-optimal bandwidth and under-smoothing. }\label{fig: 3}
\end{figure}

By default, \code{lpbwdensity()} computes the MSE-optimal bandwidth for estimating the PDF with a local quadratic regression and triangular kernel, on nineteen quantile-spaced grid points: \code{lpbwdensity(..., p=2, v=1, bwselect="mse-dpi", kernel="triangular")}. The output resembles that of \code{lpdensity()}, and provides basic information for the data and options specified, as well as a matrix with three columns: (i) \code{Grid} for grid of evaluation points, (ii) \code{B.W.} for estimated bandwidths, and (iii) \code{Eff.n} for effective sample size at each grid point given the estimated bandwidth. The following is an example with a user-chosen grid of evaluation points.
\begin{CodeChunk}
\begin{CodeInput}
R> model1bw <- lpbwdensity(data, grid = seq(0, 4, 0.5))
R> summary(model1bw)
\end{CodeInput}
\begin{CodeOutput}
Call: lpbwdensity

Sample size                           (n=)    2000
Polynomial order for point estimation (p=)    2
Density function estimated            (v=)    1
Kernel function                               triangular
Bandwidth selection method                    mse-dpi

================================
Index     Grid      B.W.   Eff.n
================================
1       0.0000    0.4064     287
2       0.5000    0.6266     933
3       1.0000    0.4721     872
4       1.5000    0.6474    1048
5       2.0000    1.0662    1216
--------------------------------
6       2.5000    0.5835     385
7       3.0000    0.5991     175
8       3.5000    0.6458      80
9       4.0000    0.6170      22
================================
\end{CodeOutput}
\end{CodeChunk}
The estimated bandwidths from this function can be used as input for \code{lpdensity()}, but constructing bandwidths in a separate step is redundant: bandwidth selection can be specified directly through the option \code{bwselect} in \code{lpdensity()}. For example, the following first computes the IMSE-optimal bandwidth and then estimates the density function:
\begin{CodeChunk}
\begin{CodeInput}
R> model5 <- lpdensity(data, grid = seq(0, 4, 0.5), bwselect="imse-dpi")
R> summary(model5)
\end{CodeInput}
\end{CodeChunk}
It may be helpful to estimate bandwidths in a separate, first step so that they can be modified prior to estimation and inference (e.g., to implement \textit{ad hoc} undersmoothing). To show this procedure, we reproduce panel (b) of Figure \ref{fig: 1} with the estimated IMSE-optimal bandwidth as well as \textit{ad hoc} under-smoothing (where the IMSE-optimal bandwidth is divided by 2). See the following code and Figure \ref{fig: 3}.
\begin{CodeChunk}
\begin{CodeInput}
R> model6bwIMSE <- lpbwdensity(data, grid = seq(0, 4, 0.05), 
+    bwselect = "imse-dpi")
R> model6 <- lpdensity(data, grid = seq(0, 4, 0.05), 
+    bw = model6bwIMSE$BW[, "bw"])
R> plot(model6) + theme(legend.position = "none")
R> model7 <- lpdensity(data, grid = seq(0, 4, 0.05), 
+    bw = model6bwIMSE$BW[, "bw"] / 2)
R> plot(model7) + theme(legend.position = "none")
\end{CodeInput}
\end{CodeChunk}

To prevent the estimated bandwidth from being too small, the default implementation in the \code{lpbwdensity()} function requires the local neighborhood around the evaluation point to contain at least $20+p+1$ (unique) observations. If the resulting neighborhood is not large enough, then the bandwidth is enlarged until the minimum number of observations is met. The default values can be changed through the options \code{nLocalMin}, controlling the minimum number of observations in each local neighborhood, and \code{nUniqueMin}, controlling the minimum number of unique observations in each local neighborhood. This minimal local sample size checking feature can be turned off by setting \code{regularize=FALSE}. Finally, the package \pkg{lpdensity} also includes checks and adjustments for repeated observations of the variable $X$ in the data. This feature can be turned off by setting \texttt{massPoints=FALSE}.

%%%%%%%%%%%%%%%%%%%%%%%%%%%%%%%%%%%%%%%%%%%%%%%%%
%% SECTION: Simulation and Comparison    	   %%
%%%%%%%%%%%%%%%%%%%%%%%%%%%%%%%%%%%%%%%%%%%%%%%%%
\section[Simulation evidence and comparison with other R packages]{Simulation evidence and comparison with other \proglang{R} packages}\label{sec:simulation}

We illustrate the finite-sample performance of our \pkg{lpdensity} package in a simulation study, and compare it with other \proglang{R} packages implementing kernel-based density estimation procedures. The functions/packages we consider are: \code{bkde()} and \code{locpoly()} in the \pkg{KernSmooth} package \citep{Wand-Ripley_2015_KernSmooth}, \code{kdde()} and \code{kde()} in the \pkg{ks} package \citep{Duong_2007_JSS,Duong_ks}, \code{npudens()} and \code{npuniden.boundary()} in the \pkg{np} package \citep{Hayfield-Racine_2008_JSS,Racine-np}, \code{kdrobust()} in the \pkg{nprobust} package \citep*{Calonico-Cattaneo-Farrell_2019_JSS,Calonico-nprobust}, \code{plugin.density()} in the \pkg{plugdensity} package \citep{Herrmann-Maechler_2011_plugdensity}, as well as the built-in density estimator \code{stats::density()}.

Table \ref{table:comparison} provides a brief summary of their main features. First, three packages offer valid density estimates at (or near) boundaries, including \pkg{KernSmooth}, \pkg{np} and our \pkg{lpdensity}. However, only \pkg{KernSmooth} and \pkg{lpdensity} provide automatic boundary carpentry, while \pkg{np} requires specifying boundary kernels. Second, only two packages, \pkg{KernSmooth} and \pkg{lpdensity}, support higher-order bias reduction. Third, statistical inference is available in \pkg{np}, \pkg{nprobust}, and \pkg{lpdensity}. However, among these three packages, only \pkg{nprobust} and \pkg{lpdensity} account for the possibly leading smoothing bias when constructing test statistics/confidence intervals using (I)MSE-optimal bandwidths, and \pkg{nprobust} is not valid at or near boundary points. In addition, our \pkg{lpdensity} package is the only one that supports constructing uniform confidence bands. Fourth, only three packages, \pkg{KernSmooth}, \pkg{ks} and \pkg{lpdensity}, offer density derivative estimation. In summary, the \pkg{lpdensity} package provides valid density and derivatives estimation for both interior and boundary evaluation points, allows higher-order bias reduction through the use of higher-order local polynomial approximations, and offers several (I)MSE-optimal bandwidth selection methods. For statistical inference, \pkg{lpdensity} takes into account the possibly leading smoothing bias, and hence delivers (asymptotically) valid testings and confidence intervals/bands, both pointwise and uniformly over evaluation points.

\begin{table}[!t]
{\centering
\renewcommand{\arraystretch}{1.2}

\resizebox{1\columnwidth}{!}{
\begin{tabular}{lrrrrrrcrrrrrr}
\hline\hline
\multicolumn{1}{l}{\bfseries }&\multicolumn{6}{c}{\bfseries Truncated Normal}&\multicolumn{1}{c}{\bfseries }&\multicolumn{6}{c}{\bfseries Exponential}\tabularnewline
\cline{2-7} \cline{9-14}
\multicolumn{1}{l}{}&\multicolumn{1}{c}{$h$}&\multicolumn{1}{c}{Bias}&\multicolumn{1}{c}{SD}&\multicolumn{1}{c}{RMSE}&\multicolumn{1}{c}{EC}&\multicolumn{1}{c}{IL}&\multicolumn{1}{c}{}&\multicolumn{1}{c}{$h$}&\multicolumn{1}{c}{Bias}&\multicolumn{1}{c}{SD}&\multicolumn{1}{c}{RMSE}&\multicolumn{1}{c}{EC}&\multicolumn{1}{c}{IL}\tabularnewline
\hline
\multicolumn{2}{l}{{\bfseries Interior ($x=1.5$)}}&&&&&&&&&&&&\tabularnewline
~~\texttt{bkde}&$$&$0.008$&$0.019$&$0.020$&$$&$$&&$$&$0.009$&$0.013$&$0.016$&$$&$$\tabularnewline
~~\texttt{locpoly}&$0.172$&$0.004$&$0.023$&$0.023$&$$&$$&&$0.100$&$0.001$&$0.024$&$0.024$&$$&$$\tabularnewline
~~\texttt{kdde}&$0.172$&$0.004$&$0.023$&$0.023$&$$&$$&&$0.100$&$0.001$&$0.024$&$0.024$&$$&$$\tabularnewline
~~\texttt{kde}&$0.172$&$0.004$&$0.023$&$0.023$&$$&$$&&$0.100$&$0.001$&$0.024$&$0.024$&$$&$$\tabularnewline
~~\texttt{npudens}&$0.102$&$0.000$&$0.035$&$0.035$&$0.964$&$0.140$&&$0.143$&$0.003$&$0.023$&$0.023$&$0.949$&$0.089$\tabularnewline
~~\texttt{npuniden.boundary}&$0.231$&$0.008$&$0.023$&$0.024$&$0.948$&$0.091$&&$0.147$&$0.003$&$0.021$&$0.021$&$0.962$&$0.084$\tabularnewline
~~\texttt{kdrobust}&$0.609$&$0.011$&$0.016$&$0.019$&$0.936$&$0.081$&&$0.633$&$0.009$&$0.013$&$0.016$&$0.941$&$0.063$\tabularnewline
~~\texttt{plugin.density}&$0.144$&$0.003$&$0.026$&$0.026$&$$&$$&&$0.071$&$0.000$&$0.029$&$0.029$&$$&$$\tabularnewline
~~\texttt{density}&$0.179$&$0.004$&$0.022$&$0.023$&$$&$$&&$0.185$&$0.004$&$0.017$&$0.018$&$$&$$\tabularnewline
~~\texttt{lpdensity}($h_{\mathtt{MSE}}$)&$0.785$&$0.008$&$0.021$&$0.022$&$0.957$&$0.102$&&$0.680$&$0.006$&$0.015$&$0.017$&$0.949$&$0.083$\tabularnewline
~~\texttt{lpdensity}($h_{\mathtt{IMSE}}$)&$0.623$&$0.007$&$0.019$&$0.020$&$0.947$&$0.112$&&$0.687$&$0.007$&$0.014$&$0.016$&$0.948$&$0.083$\tabularnewline
\hline
\multicolumn{2}{l}{{\bfseries Near boundary ($x=0.2$)}}&&&&&&&&&&&&\tabularnewline
~~\texttt{locpoly}&$0.172$&$0.033$&$0.024$&$0.041$&$$&$$&&$0.100$&$0.020$&$0.045$&$0.049$&$$&$$\tabularnewline
~~\texttt{npuniden.boundary}&$0.230$&$0.019$&$0.026$&$0.032$&$0.877$&$0.099$&&$0.147$&$0.018$&$0.043$&$0.046$&$0.931$&$0.172$\tabularnewline
~~\texttt{lpdensity}($h_{\mathtt{MSE}}$)&$1.149$&$0.022$&$0.044$&$0.049$&$0.948$&$0.118$&&$0.903$&$0.009$&$0.045$&$0.046$&$0.938$&$0.153$\tabularnewline
~~\texttt{lpdensity}($h_{\mathtt{IMSE}}$)&$0.621$&$0.001$&$0.030$&$0.030$&$0.950$&$0.117$&&$0.687$&$0.001$&$0.040$&$0.040$&$0.944$&$0.156$\tabularnewline
\hline
\multicolumn{1}{l}{{\bfseries Boundary ($x=0$)}}&&&&&&&&&&&&\tabularnewline
~~\texttt{locpoly}&$0.172$&$0.139$&$0.016$&$0.140$&$$&$$&&$0.100$&$0.548$&$0.034$&$0.549$&$$&$$\tabularnewline
~~\texttt{npuniden.boundary}&$0.230$&$0.054$&$0.033$&$0.063$&$0.506$&$0.117$&&$0.147$&$0.091$&$0.083$&$0.124$&$0.548$&$0.242$\tabularnewline
~~\texttt{lpdensity}($h_{\mathtt{MSE}}$)&$0.686$&$0.010$&$0.058$&$0.059$&$0.944$&$0.348$&&$0.807$&$0.045$&$0.087$&$0.098$&$0.932$&$0.511$\tabularnewline
~~\texttt{lpdensity}($h_{\mathtt{IMSE}}$)&$0.621$&$0.007$&$0.055$&$0.055$&$0.955$&$0.343$&&$0.687$&$0.026$&$0.082$&$0.086$&$0.952$&$0.514$\tabularnewline
\hline
\end{tabular}
}

\caption{Simulation results.}\label{table:simulations} \medskip
}

\footnotesize{Notes. Empty cells correspond to features that are not readily available without modifying the source code. For the case of ``Near boundary'' and ``Boundary'' we only consider software packages/functions that are valid for those cases. Default options for each package/function are used whenever possible. Results are based on 2,000 simulations with a sample size of 1,000. Column ``Truncated Normal'': the $\mathcal{N}(1,1)$ distribution truncated from below at 0. Column ``Exponential'': the exponential distribution with a scale parameter 1.}
\end{table}

We now describe our simulation design. The data consists of a random sample of size $n=1,000$, generated either from the normal distribution $\mathcal{N}(1,1)$ truncated below at $0$ (column ``Truncated Normal''), or the exponential distribution with a scale parameter of 1 (column ``Exponential''). We consider the estimation of the PDF at three evaluation points: $x=1.5$, $x=0.2$ and $x=0$, corresponding to interior, near boundary and boundary regions, respectively. We employ $2,000$ Monte Carlo repetitions. For the point estimate, we report its bias (column ``Bias''), standard deviation (column ``SD'') and root mean squared error (column ``RMSE''). Whenever available, we also report the empirical coverage probability of a nominal 95\% confidence interval (column ``EC'') as well as its average length (column ``IL'').

Simulation results are reported in Table \ref{table:simulations}. At the interior evaluation point, all procedures perform similarly in terms of RMSE and empirical coverage. Point estimates obtained using \code{lpdensity()} have relatively small RMSEs, and the corresponding RBC confidence intervals exhibit satisfactory coverage properties. When the evaluation point is close to or exactly at the boundary, most packages/functions are no longer valid, and hence we only report simulation results for \code{locpoly()}, \code{npuniden.boundary()}, and \code{lpdensity()}. In such cases, \code{lpdensity()} delivers points estimates and confidence intervals with excellent finite-sample performance.

%%%%%%%%%%%%%%%%%%%%%%%%%%%%%%%%%%%%%%%%%%%%%%%%%
%% SECTION: Conclusion                         %%
%%%%%%%%%%%%%%%%%%%%%%%%%%%%%%%%%%%%%%%%%%%%%%%%%
\section{Conclusion}\label{sec:conclusion}

We gave an introduction to the general purpose software package \pkg{lpdensity}, which offers local polynomial regression based estimation and inference procedures for a cumulative distribution function, probability density function, and higher-order derivatives thereof. This package is available in both \proglang{R} and \proglang{Stata} statistical platforms, and further details can be found at \url{https://nppackages.github.io/lpdensity/}.

%%%%%%%%%%%%%%%%%%%%%%%%%%%%%%%%%%%%%%%%%%%%%%%%%
%% SECTION: Acknowledgments                    %%
%%%%%%%%%%%%%%%%%%%%%%%%%%%%%%%%%%%%%%%%%%%%%%%%%
\section*{Acknowledgments}

We thank Sebastian Calonico, David Drukker, Yingjie Feng, the editor, and two anonymous reviewers for thoughtful comments on our software implementation and article. We are also grateful to many users who provided valuable feedback. Cattaneo gratefully acknowledges financial support from the National Science Foundation (SES-1459931 and SES-1947805). Jansson gratefully acknowledges financial support from the National Science Foundation (SES-1459967 and SES-1947662) and the research support of CREATES.

%% -- Bibliography -------------------------------------------------------------
%% - References need to be provided in a .bib BibTeX database.
%% - All references should be made with \cite, \citet, \citep, \citealp etc.
%%   (and never hard-coded). See the FAQ for details.
%% - JSS-specific markup (\proglang, \pkg, \code) should be used in the .bib.
%% - Titles in the .bib should be in title case.
%% - DOIs should be included where available.

\bibliography{Cattaneo-Jansson-Ma_2021_JSS}
\clearpage

%% -- Appendix (if any) --------------------------------------------------------
%% - After the bibliography with page break.
%% - With proper section titles and _not_ just "Appendix".

\begin{appendix}

%%%%%%%%%%%%%%%%%%%%%%%%%%%%%%%%%%%%%%%%%%%%%%%%%
%% Appendix Bandwidth Selectors                %%
%%%%%%%%%%%%%%%%%%%%%%%%%%%%%%%%%%%%%%%%%%%%%%%%%
\section{Details on bandwidth selection}\label{appendix:details}\setcounter{equation}{0}

We provide more methodological details on bandwidth selectors implemented thorough the function \code{lpbwdensity()}. We continue to focus on the case of $1\leq\nu\leq p$, and therefore do not discuss bandwidth selection for CDF estimation. See CJM for details. 

\subsection{Rule-of-thumb bandwidths}
Recall that, in the definition of $\mathsf{Var}[\hat{F}_p^{(\nu)}(x)]$ and $\mathsf{Bias}[\hat{F}_p^{(\nu)}(x)]$, we introduced pre-asymptotic quantities $\mathsf{V}_{\nu,p}(x)$, $\mathsf{B}_{1,\nu,p}(x)$ and $\mathsf{B}_{2,\nu,p}(x)$. For the rule-of-thumb bandwidth selectors, we consider a normal reference model, hence all evaluation points are interior. Then, those quantities have well-defined limits, which can be computed using features of the underlying distribution (such as normal densities and higher-order derivatives), $p$, $\nu$, and the kernel function. We denote the rule-of-thumb bandwidth by $\hat{h}_{\mathtt{ROT},p}$. An integrated version can be constructed accordingly, and is denoted by $\hat{h}_{\mathtt{IROT},p}$. 

Given $x$, $p$ and $\nu$, the rate at which the MSE-optimal bandwidth $h_{\mathtt{MSE}}$ shrinks to zero depends on whether $p-\nu$ is even or odd, and whether $x$ is interior or boundary. This is summarized in panel (a) of Table \ref{table: 1}. We also include the rate at which the rule-of-thumb bandwidths shrinks in panel (b). (The notation $\hat{h}\asymp_{\P} n^{-1/\gamma}$ indicates that both $n^{1/\gamma}\hat{h}$ and $n^{-1/\gamma}\hat{h}^{-1}$ are bounded in probability.) Note that the (I)ROT-optimal bandwidths have the correct rate of convergence, except when $p-\nu$ is even and $x$ is near boundary.  

\subsection{(I)MSE-optimal bandwidths}

We now discuss some implementation details of the MSE-optimal bandwidth, which will also apply to the construction of the IMSE-optimal bandwidth. First, the unknown higher-order derivatives $F^{(p+1)}(x)$ and $F^{(p+2)}(x)$ are replaced by consistent estimates, $\hat{F}_{p+2}^{(p+1)}(x;\hat{h}_{\mathtt{IROT},p+1,p+2})$ and $\hat{F}_{p+3}^{(p+2)}(x;\hat{h}_{\mathtt{IROT},p+2,p+3})$, respectively. Here we augment the subscript of bandwidths with one additional argument, since the bandwidth depends on both the polynomial order as well as the order of derivative. For example, $\hat{h}_{\mathtt{IROT},p+1,p+2}$ is an estimated bandwidth using a normal reference model, which is IMSE-optimal for a local polynomial regression of order $p+2$ when estimating the $(p+1)$-th derivative of $F(x)$. 

The next step is to construct the pre-asymptotic quantities $\mathsf{V}_{\nu,p}(x)$, $\mathsf{B}_{1,\nu,p}(x)$ and $\mathsf{B}_{2,\nu,p}(x)$, which require a preliminary bandwidth. We use $\hat{h}_{\mathtt{IROT},1,2}$, so those quantities are $\mathsf{V}_{\nu,p}(x;\hat{h}_{\mathtt{IROT},1,2})$, $\mathsf{B}_{1,\nu,p}(x;\hat{h}_{\mathtt{IROT},1,2})$ and $\mathsf{B}_{2,\nu,p}(x;\hat{h}_{\mathtt{IROT},1,2})$. Then, the MSE-optimal bandwidth is
\[
\hat{h}_{\mathtt{MSE},p} = \argmin_{h>0} \left\{\widehat{\mathsf{Var}}[\hat{F}_p^{(\nu)}(x)] + \widehat{\mathsf{Bias}}[\hat{F}_p^{(\nu)}(x)]^2\right\},
\]
with $\widehat{\mathsf{Var}}[\hat{F}_p^{(\nu)}(x)] = \frac{1}{nh^{2\nu-1}}\mathsf{V}_{\nu,p}(x;\hat{h}_{\mathtt{IROT},1,2})$ and
\begin{align*}
	\widehat{\mathsf{Bias}}[\hat{F}_p^{(\nu)}(x)] = h^{p-\nu+1} &\Big[\hat{F}_{p+2}^{(p+1)}(x;\hat{h}_{\mathtt{IROT},p+1,p+2})\mathsf{B}_{1,\nu,p}(x;\hat{h}_{\mathtt{IROT},1,2})\\
	                                                            &\qquad + h \cdot \hat{F}_{p+3}^{(p+2)}(x;\hat{h}_{\mathtt{IROT},p+2,p+3})\mathsf{B}_{2,\nu,p}(x;\hat{h}_{\mathtt{IROT},1,2})\Big].
\end{align*}

Under regularity conditions, it can be shown that $\hat{h}_{\mathtt{MSE},p}$ is rate consistent (see panel (c) of Table \ref{table: 1}). Under the assumption that either (i) $x$ is near boundary, or (ii) $p-\nu$ is odd, it is possible to show a stronger result: $\hat{h}_{\mathtt{MSE},p}/{h}_{\mathtt{MSE},p}\overset{\P}{\to} 1$, so that the MSE-optimal bandwidth selector is consistent both in \emph{rate} and \emph{constant}. What happens for interior $x$ with $p-\nu$ even? In this case $\mathsf{B}_{1,\nu,p}(x;\hat{h}_{\mathtt{IROT},1,2})\overset{\P}{\to} 0$, and $\mathsf{B}_{2,\nu,p}(x;\hat{h}_{\mathtt{IROT},1,2})$ captures only part of the leading bias. As a result, $\hat{h}_{\mathtt{MSE},p}$ has the correct rate of convergence, but is not consistent for ${h}_{\mathtt{MSE},p}$ in the strong sense. 

\begin{table}[t!]
\centering
\subfloat[$h_{\mathtt{MSE},p}\asymp n^{-1/\gamma}$]{\resizebox{0.4\columnwidth}{!}{
\begin{tabular}{rcc}
\hline 
& $x$ interior & $x$ boundary \\ \hline
$p-\nu$ odd    & $\gamma=2p+1$       & $\gamma=2p+1$ \\
$p-\nu$ even   & $\gamma=2p+3$       & $\gamma=2p+1$ \\ \hline
\end{tabular}
}}\subfloat[$\hat{h}_{\mathtt{ROT},p}\asymp_{\P} n^{-1/\gamma}$ and $\hat{h}_{\mathtt{IROT},p}\asymp_{\P} n^{-1/\gamma}$]{\resizebox{0.4\columnwidth}{!}{
\begin{tabular}{rcc}
\hline 
& $x$ interior & $x$ boundary \\ \hline
$p-\nu$ odd    & $\gamma=2p+1$ & $\gamma=2p+1$ \\
$p-\nu$ even   & $\gamma=2p+3$ & $\gamma=2p+3$ \\ \hline
\end{tabular}
}}
	
\subfloat[$\hat{h}_{\mathtt{MSE},p}\asymp_{\P} n^{-1/\gamma}$]{\resizebox{0.4\columnwidth}{!}{
\begin{tabular}{rcc}
\hline 
& $x$ interior & $x$ boundary \\ \hline
$p-\nu$ odd    & $\gamma=2p+1$ & $\gamma=2p+1$ \\
$p-\nu$ even   & $\gamma=2p+3$ & $\gamma=2p+1$ \\ \hline
\end{tabular}
}}
\caption{Bandwidths rates for $1\leq \nu\leq p$. }\label{table: 1}
\end{table}

%%%%%%%%%%%%%%%%%%%%%%%%%%%%%%%%%%%%%%%%%%%%%%%%%
%% Appendix Bandwidth Selectors                %%
%%%%%%%%%%%%%%%%%%%%%%%%%%%%%%%%%%%%%%%%%%%%%%%%%
\section[Stata Implementation]{\proglang{Stata} Implementation}\label{appendix:stata}

We discuss the \proglang{Stata} implementation of our \pkg{lpdensity} package, which offers two commands, \code{lpdensity} for estimation of and inference on the CDF, PDF, and their higher-order derivatives, and \code{lpbwdensity} for data-driven bandwidth selection. The plotting features employ the built-in command \code{twoway}.

The command \code{lpdensity} provides point estimation and robust confidence intervals/bands employing the local polynomial density estimator, given a grid of points and a bandwidth choice. We generate 2,000 observations from the normal distribution $\mathcal{N}(1,1)$ truncated below at 0. Although the same seed, 42, as in \proglang{R} is used, observations generated in \proglang{Stata} are generally different due to the different random number generators used by the statistical platforms. 

The following command estimates the density function (\code{v(1)}, the default) with fixed bandwidth \code{bw(0.5)} over the grid of evaluation points $0, 0.5, \cdots, 4$, using a local quadratic  approximation (\code{p(2)}, the default) to the empirical distribution function. Robust bias-corrected confidence intervals over the grid are computed using a local cubic approximation (\code{q(3)}, the default). 
\begin{CodeChunk}
\begin{CodeInput}
. set seed 42
. set obs 4000
. gen data = rnormal(1, 1)
. drop if data <= 0 
. drop if _n > 2000 
. gen grid = -0.5 + 0.5 * _n if _n <= 9
. lpdensity data, grid(grid) bw(0.5)
\end{CodeInput}
\begin{CodeOutput}
Local Polynomial Density Estimation and Inference.

 Sample size                              (n=)               2000
 Polynomial order for point estimation    (p=)                  2
 Density function estimated               (v=)                  1
 Polynomial order for confidence interval (q=)                  3
 Kernel function                                       triangular
 Bandwidth selection method                               mse-dpi

------------------------------------------------------------------------
                                     Point      Std.         Robust B.C.
 Index    Grid      B.W.   Eff.n      Est.     Error            95% C.I.
------------------------------------------------------------------------
   1    0.0000    0.5000     366    0.2815    0.0403    0.1064    0.3547
   2    0.5000    0.5000     814    0.4230    0.0153    0.3899    0.4829
   3    1.0000    0.5000     897    0.4680    0.0158    0.4455    0.5414
   4    1.5000    0.5000     834    0.4056    0.0146    0.3594    0.4486
   5    2.0000    0.5000     607    0.3209    0.0143    0.2981    0.3810
------------------------------------------------------------------------
   6    2.5000    0.5000     307    0.1406    0.0100    0.1065    0.1632
   7    3.0000    0.5000     117    0.0516    0.0061    0.0303    0.0642
   8    3.5000    0.5000      43    0.0199    0.0038    0.0070    0.0292
   9    4.0000    0.5000      12    0.0130    0.0077   -0.0006    0.0432
------------------------------------------------------------------------
\end{CodeOutput}
\end{CodeChunk}
Coverage of the robust confidence interval can be specified through \code{level()}. For example, to report nominal 99\% confidence intervals, one can use 
\begin{CodeChunk}
\begin{CodeInput}
. lpdensity data, grid(grid) bw(0.5) level(99)
\end{CodeInput}
\end{CodeChunk}
When the argument \code{grid()} is suppressed, the evaluation points will be the $0.05, 0.1, \cdots,0.9, 0.95$ quantiles computed from the data. Conventional inference results (i.e., without robust bias correction) can be obtained by setting \code{q()} to be the same as \code{p()}. 
\begin{CodeChunk}
\begin{CodeInput}
. lpdensity data, bw(0.5)
. lpdensity data, bw(0.5) q(2)
\end{CodeInput}
\end{CodeChunk}

In \proglang{Stata}, graphical illustration of the estimates can be obtained using the option \code{plot}. The following plots the estimated density function on a fine grid, which resembles Panel (b) of Figure \ref{fig: 1}.
\begin{CodeChunk}
\begin{CodeInput}
. capture drop grid
. gen grid = -0.05 + 0.05 * _n if _n <= 81
. lpdensity data, grid(grid) bw(0.5) plot
\end{CodeInput}
\end{CodeChunk}
The same figure can be produced by first storing estimation results and then calling the \code{twoway} command directly. For example, 
\begin{CodeChunk}
\begin{CodeInput}
. lpdensity data, grid(grid) bw(0.5) genvars(lpdTemp)
. twoway                                                               ///
>   (rarea lpdTemp_CI_l lpdTemp_CI_r lpdTemp_grid, sort color(red%30)) ///
>   (line lpdTemp_f_p lpdTemp_grid,                                    /// 
>     sort lcolor(red) lwidth("medthin") lpattern(solid)),             ///
>   legend(off) title("lpdensity (p=2, q=3, v=1)", color(gs0))         ///
>   xtitle("data") ytitle("")
. drop lpdTemp_*
\end{CodeInput}
\end{CodeChunk}
To further illustrate, the following generates analogues of Panel (c) and (d) of Figure \ref{fig: 2}. 
\begin{CodeChunk}
\begin{CodeInput}
. lpdensity data, grid(grid) bw(0.5) plot histogram
. lpdensity data, grid(grid) bw(0.5) plot ciuniform level(90)
\end{CodeInput}
\end{CodeChunk}

Before closing this appendix, we illustrate the bandwidth selector \code{lpbwdensity}. By default, this command computes the MSE-optimal bandwidth for estimating the PDF with a local quadratic regression and triangular kernel: 
\begin{CodeChunk}
\begin{CodeInput}
. capture drop grid
. gen grid = -0.5 + 0.5 * _n if _n <= 9
. lpbwdensity data, grid(grid) 
\end{CodeInput}
\begin{CodeOutput}
Bandwidth Selection for Local Polynomial Density Estimation.

 Sample size                              (n=)            2000
 Polynomial order for point estimation    (p=)               2
 Density function estimated               (v=)               1
 Kernel function                                    triangular
 Bandwidth selection method                            mse-dpi

--------------------------------
 Index    Grid      B.W.   Eff.n
--------------------------------
   1    0.0000    0.3812     258
   2    0.5000    0.6167     947
   3    1.0000    0.5254     946
   4    1.5000    0.7212    1168
   5    2.0000    0.5599     667
--------------------------------
   6    2.5000    0.4835     298
   7    3.0000    0.4925     114
   8    3.5000    1.1727     183
   9    4.0000    0.7140      22
--------------------------------
\end{CodeOutput}
\end{CodeChunk}
Finally, the following computes the IMSE-optimal bandwidth for density estimation. 
\begin{CodeChunk}
\begin{CodeInput}
. lpbwdensity data, grid(grid) bwselect(imse-dpi)
\end{CodeInput}
\end{CodeChunk}
\end{appendix}
\clearpage

\end{document}